\documentclass[fleqn,usenatbib]{mnras}

\usepackage[T1]{fontenc}
\usepackage{ae,aecompl}
\usepackage{graphicx}	% Including figure files
\usepackage{amsmath}	% Advanced maths commands
\usepackage{amssymb}	% Extra maths symbols
\usepackage{color}
\usepackage{soul}

\newcommand{\lsim }{{\lower0.8ex\hbox{$\buildrel <\over\sim$}}}
\newcommand{\gsim }{{\lower0.8ex\hbox{$\buildrel >\over\sim$}}}

\def\apj{ApJ}
\def\apjl{ApJL}
\def\aap{A\&A}
\def\mnras{MNRAS}
\def\pasp{PASP}

\def\aj{AJ}

\def\apjs{ApJ Supp}
\def\nar{New A Rev.} 

\def\nat{Natur}

\def\nustar{\emph{NuSTAR}}
\def\ATCA{ATCA}
\def\Einstein{\emph{Einstein}}
\def\Einsteinipc{\emph{Einstein/IPC}}
\def\Einsteinhri{\emph{Einstein/HRI}}
\def\chandra{\emph{Chandra}}
\def\chandraacis{\emph{Chandra/ACIS}}
\def\chandrahrc{\emph{Chandra/HRC}}

\def\rosat{\emph{ROSAT}}
\def\rosathri{\emph{ROSAT/HRI}}
\def\Swiftxrt{\emph{Swift/XRT}}

\def\Swift{\emph{Swift}}
\def\HST{${\it HST}$}

\def\simge{\mathrel{%
  \rlap{\raise 0.511ex \hbox{$>$}}{\lower 0.511ex \hbox{$\sim$}}}}
\def\simle{\mathrel{
  \rlap{\raise 0.511ex \hbox{$<$}}{\lower 0.511ex \hbox{$\sim$}}}}

\newcommand{\Msun}{\ifmmode {M_{\odot}}\else${M_{\odot}}$\fi}
\newcommand{\Lsun}{\ifmmode {L_{\odot}}\else${L_{\odot}}$\fi}
\newcommand{\Rsun}{\ifmmode {R_{\odot}}\else${R_{\odot}}$\fi}

\title[Ultracompact XRB 47 Tuc X9]{The ultracompact nature of the black hole candidate X-ray binary 47 Tuc X9}

\author[Bahramian et al.]{Arash Bahramian$^{1,2}$\thanks{E-mail: bahramian@pa.msu.edu}, Craig O. Heinke$^{1}$, Vlad Tudor$^{3}$, James C.A. Miller-Jones$^{3}$,
\newauthor Slavko Bogdanov$^{4}$, Thomas J. Maccarone$^{5}$, Christian Knigge$^{6}$, Gregory R. Sivakoff$^{1}$, 
\newauthor Laura Chomiuk$^{2}$, Jay Strader$^{2}$, Javier A. Garcia$^{7}$, Timothy Kallman$^{8}$
\\
$^{1}$ Department of Physics, CCIS 4-183, University of Alberta, Edmonton, AB T6G 2E1, Canada\\
$^{2}$ Department of Physics and Astronomy, Michigan State University, East Lansing, MI, USA\\
$^{3}$ International Centre for Radio Astronomy Research - Curtin University, GPO Box U1987, Perth, WA 6845, Australia\\
$^{4}$ Columbia Astrophysics Laboratory, Columbia University, New York, NY, USA\\
$^{5}$ Department of Physics, Texas Tech University, Lubbock, TX, USA\\
$^{6}$ School of Physics and Astronomy, University of Southampton, Southampton, UK\\
$^{7}$ Harvard-Smithsonian Center for Astrophysics, Cambridge, MA, USA\\
$^{8}$ NASA Goddard Space Flight Center, Greenbelt, MD, USA\\
}

\pubyear{2016}

\begin{document}
\label{firstpage}
\pagerange{\pageref{firstpage}--\pageref{lastpage}}
\maketitle

% Abstract of the paper
\begin{abstract}
47 Tuc X9 is a low mass X-ray binary (LMXB) in the globular cluster 47 Tucanae, and was previously thought to be a cataclysmic variable. However, \citet{Miller-Jones15} recently identified a radio counterpart to X9 (inferring a radio/X-ray luminosity ratio consistent with black hole LMXBs), and suggested that the donor star might be a white dwarf.  We report simultaneous observations of X9 performed by \chandra, \nustar\ and Australia Telescope Compact Array. We find a clear 28.18$\pm0.02$-min periodic modulation in the \chandra\ data, which we identify as the orbital period, confirming this system as an ultracompact X-ray binary. Our X-ray spectral fitting provides evidence for photoionized gas having a high oxygen abundance in this system, which indicates a C/O white dwarf donor. We also identify reflection features in the hard X-ray spectrum, making X9 the faintest LMXB to show X-ray reflection. We detect a $\sim$ 6.8-d modulation in the X-ray brightness by a factor of 10, in archival \chandra, \Swift, and \rosat\ data. The simultaneous radio/X-ray flux ratio is consistent with either a black hole primary or a neutron star primary, if the neutron star is a transitional millisecond pulsar. Considering the measured orbital period (with other evidence of a white dwarf donor), and the lack of transitional millisecond pulsar features in the X-ray light curve, we suggest that this could be the first ultracompact black hole X-ray binary identified in our Galaxy.
\end{abstract}

\begin{keywords}
accretion, accretion discs, stars: neutron, black holes, X-rays: binaries, globular clusters: individual: 47 Tuc
\end{keywords}

%%%%%%%%%%%%%%%%%%%%%%%%%%%%%%%%%%%%%%%%%%%%%%%%%%

%%%%%%%%%%%%%%%%% BODY OF PAPER %%%%%%%%%%%%%%%%%%
\section{Introduction}
Low mass X-ray binaries (LMXBs) are systems in which a compact object [neutron star (NS) or black hole (BH)] accretes matter from a low mass companion (typically a main sequence star) through Roche-lobe overflow or wind-fed accretion (from a red giant).
It has been long noticed that the population of LMXBs per unit mass in globular clusters (GCs) is orders of magnitude higher than that of the Galactic field \citep{Katz75,Verbunt87,Pooley03,Heinke03d,Jordan04}. This overabundance has been associated with the high stellar density in GCs. While most LMXBs in the Galactic field are formed through binary evolution of primordial binaries, it has been shown that in GCs, the dominant channel to form LMXBs is through encounters. These scenarios include tidal capture by an NS  or BH, exchange of a compact object into a primordial binary, or collision of a compact object with a red giant \citep{Fabian75,Sutantyo75,Hills76,Bailyn90a,Davies98,Ivanova06,Ivanova08}.

While it is generally understood that NSs are well-represented in GCs, the presence of BHs in GCs is more controversial. Hundreds to thousands of stellar mass BHs form in GCs through stellar evolution. A fraction of these BHs will have sufficient natal kicks to leave the cluster after birth while the rest will sink towards the center of the cluster and form a dense sub-cluster. It was previously thought the core-collapse of this sub-cluster will result in total evaporation from the GC and thus in most cases no BHs will be left in the cluster \citep{Kulkarni93,Sigurdsson93}. This is consistent with observations of LMXBs in GCs as all the bright (persistent or transient) LMXBs found in Galactic GCs have been shown to be NS LMXBs \citep[e.g.,][]{Verbunt06,Papitto13,Bahramian14}. For example, the radio/X-ray flux ratio of the transient in NGC 6388 strongly indicates it holds an NS accretor \citep{Bozzo11}. The NGC 2808 transient is an accreting millisecond X-ray pulsar \citep{Sanna17}. M15 X-2 produces X-ray bursts, which would not be visible from AC 211 \citep{White01}, but AC 211 also holds an NS, as shown by tomographic analysis \citep{vanZyl04}.

However, our understanding of the BH LMXB population in GCs has changed significantly over the past few years, with the discovery of highly luminous BH LMXB candidates in extragalactic GCs\footnote{ Discovery of neutron star ultra-luminous X-ray sources in recent years \citep{Bachetti14,Israel17,Israel16,Furst16} indicate that X-ray luminosity $\gsim 10^{39}$ erg/s in a single object does not guarantee a BH accretor.} \citep[e.g.,][]{Maccarone07a,Maccarone11}, as well as quiescent BH LMXB candidates in Galactic GCs \citep{Strader12,Chomiuk13,Miller-Jones15}. Additionally, on the theoretical front recent, more detailed simulations have found that a substantial fraction of BHs may remain in GCs up to the present day \citep{Sippel13,Morscher13,Morscher15,Peuten16}. The recent LIGO detection of a BH-BH merger has also rekindled interest in the formation of such binaries inside GCs \citep{Rodriguez15,Rodriguez16}.

X-ray spectroscopy can help with distinguishing between some types of faint/quiescent X-ray sources in GCs. While cataclysmic variables (CVs) --- systems with a white dwarf (WD) accreting from a main-sequence or sub-giant star --- tend to show a hard X-ray spectrum dominated by hot bremsstrahlung emission, NS LMXBs (in quiescence) often show a softer spectrum, dominated by blackbody-like emission from the NS surface \citep{Rutledge99,Rutledge02c,Grindlay01a,Heinke03d}.

X-ray spectra of quiescent LMXBs also often show a non-thermal component, fit by a power law with photon index between 1 and 2, which could be produced by continuous low-level accretion \citep[e.g.,][]{Campana98a}. While the non-thermal component in quiescent NS LMXBs may be associated either with accretion or with magnetospheric or shock emission related to an active radio pulsar wind \citep{Campana98a,Jonker04,Wijnands15}, in quiescent BH LMXBs, the X-ray emission is thought to be from a radiatively inefficient accretion flow, possibly synchrotron emission from the jet \citep{Gallo07} and/or synchrotron self-Compton emission from the jet base \citep{Plotkin15}. The non-thermal component has similar properties in NS and BH LMXBs in quiescence, so it can be difficult to distinguish between BH LMXBs and NS LMXBs lacking a strong blackbody-like component. A chief diagnostic method to find BH LMXBs has been based on the ratio of radio and X-ray luminosities of the system during the hard or the quiescent state \citep{Maccarone05a}. BH X-ray binaries (XRBs) show compact partially self-absorbed jet emission in quiescence (and in the hard state during outbursts), making them brighter in radio compared to NS LMXBs with similar X-ray luminosities \citep{Gallo03,Fender03,Migliari06}. 

Millisecond pulsars (MSPs) are old pulsars that have been spun up through accretion from a companion \citep{Alpar82}. It is thought that NS LMXBs are  progenitors of MSPs. Detection of millisecond pulsations in the LMXB SAX J1808.4-3624 \citep{Wijnands98} -- and later in other LMXBs -- clearly demonstrated this scenario. The final evolutionary link connecting LMXBs and MSPs has been found in recent years, in the form of transitional millisecond pulsars (tMSPs), in which the system moves between an accretion-powered state and a rotation-powered one on  timescales as short as weeks \citep[e.g.,][]{Archibald09,Papitto13}.

In LMXBs, if the companion is a compact, hydrogen-deficient star (e.g., a WD), then the orbital period is short ($P_{orb} < 80$ min), and the system is an ultra-compact X-ray binary (UCXB). The depletion of hydrogen in UCXBs leads to high abundances of either helium, if the donor is a low-mass He WD, or carbon and oxygen, if the donor is a more massive C/O WD \citep{Nelemans04}. In the case of UCXBs with C/O WDs, the overabundance of carbon and oxygen can produce observable diagnostics in far-UV and X-ray spectra \citep{Nelemans10b}.

47 Tucanae (47 Tuc) is a dense, massive \citep[$6.5\times10^5 M_\odot$,][]{Kimmig15} Galactic GC at a distance of 4.53 kpc\footnote{This value is derived from a compilation of 22 distance measurements, as reported by \citet{Bogdanov16}.} \citep{Bogdanov16,Woodley12}. Low extinction \citep[$E(B-V) = 0.04$,][]{Salaris07} makes this cluster easy to study. 47 Tuc harbours $\sim230$ X-ray sources with $L_X \gsim 10^{29}$ erg/s \citep{Heinke05a}, including 23 radio MSPs \citep{Freire03,Pan16}, and many CVs and chromospherically active binaries \citep{Grindlay01a,Edmonds03a,Edmonds03b,Knigge02,Knigge03,Knigge08}. As the cluster has a very high stellar encounter rate \citep{Bahramian13}, it should produce many XRBs. Indeed, five quiescent NS LMXBs, all with prominent thermal blackbody-like components in their X-ray spectra (two also have strong non-thermal components), have been identified and studied in 47 Tuc \citep{Heinke03a,Heinke05b,Heinke06a,Bogdanov16}.

X-ray emission from 47 Tuc was first noticed in observations by \Einstein, and the source was identified as 1E 0021.8-7221 \citep{Hertz83,Grindlay84}. In \HST\ observations, a variable blue star - named V1 - was discovered and identified as a likely CV and the counterpart of 1E 0021.8-7221, by \citet{Paresce92}. The relatively high X-ray luminosity, low optical/X-ray ratio, and hints of a 120-s periodicity in Einstein data \citep{Auriere89}, inspired the suggestion that V1 (1E 0021.8-7221) was an intermediate polar, where the modulation would be due to rotation of the WD \citep{Paresce92}. Later \rosat\ observations found 9 X-ray sources within the cluster, and identified X9 as the counterpart for 1E 0021.8-7221 \citep[][who argued, presciently, that X9 was likely a low-mass X-ray binary]{Verbunt98}. The association of V1 and X9 was later confirmed by \chandra\ observations \citep{Grindlay01a}. X9 showed rapid ($\sim$hours) X-ray and optical/UV variability, with a hint of a $\sim$6-hour periodicity in the UV flux \citep{Paresce92}. Time series analysis of $V$ and $I$ light curves by \citet{Edmonds03b} indicated a marginal 3.5 hour (or 7 hour, if due to ellipsoidal variations) periodicity. \HST\ far-UV-spectroscopy of X9 showed strong, double-peaked C IV emission lines, confirming the presence of an accretion disk and/or disk wind in the system \citep{Knigge08}. 

X9 is the brightest ($\gsim 10^{33}$ erg/s) X-ray source in the core of 47 Tuc with a hard spectrum \citep{Grindlay01a,Heinke05a}. In both \Einstein\ and \rosat\ observations, X9 has shown clear variability. The first \chandra\ study of X9 (aka W42, as designated by \citealt{Grindlay01a}) in 2000 suggested the presence of a 218 s modulation in the light curve, which, if verified, would mark X9 as an intermediate polar CV \citep{Grindlay01a}. Spectral analyses of the  \chandra\ X-ray data showed a complex spectrum with a very hard non-thermal component and extremely strong oxygen spectral lines \citep{Heinke05a}. 

Recently \citet{Miller-Jones15} reported on deep radio observations of 47 Tuc, in which they detected a radio counterpart for X9. They reported that the ratio of radio/X-ray luminosity for X9 is strongly inconsistent with a CV, and more consistent with a BH LMXB, or possibly a tMSP (deemed unlikely from the X-ray spectral and variability properties). Based on the brightness (and thus accretion rate) and evidence against strong H$\alpha$ emission from \emph{HST} narrow-band photometry, they suggested the system might be a UCXB. They further inferred an orbital period of $\sim 25$ min, from a correlation between quiescent X-ray luminosity and time-averaged mass transfer rate from the companion, and using calculations of the time-averaged mass transfer rate at different orbital periods in UCXBs. Note that this was not based on detection of a periodic signal.

In this work we study the X-ray properties of X9, as observed by \chandra\ and \nustar, and the radio/X-ray flux ratio from simultaneous radio observations. We describe the data used in this work, along with a description of reduction and analysis methods, in Section~\ref{sec_data}. In Section~\ref{sec_results}, we present results and in Section~\ref{sec_disc}, we will discuss the implications of these findings. 

\section{Data and Analysis}\label{sec_data}
We obtained simultaneous radio and X-ray observations of X9 with the Australia Telescope Compact Array (\ATCA), \chandra\ using the ACIS-S detector, and \nustar\ on February 2nd, 2015 (MJD = 57055; \ATCA\ from 05:00 to 17:00 UT on Feb 2nd, \chandra\ from 14:25 on Feb. 2nd,  to 03:20 on Feb. 3rd, and \nustar\ from  11:51 on Feb. 2nd to 00:46 on Feb. 5th, all UT).  We also analyzed \chandra\ observations from late 2014 \citep[described in][]{Bogdanov16} and  archival \chandraacis, \chandrahrc, \Swiftxrt, \rosathri, and \Einstein\ data to study the variability of X9 over long time scales ($\sim$ days -- years).
 
\subsection{Chandra/ACIS}
Archival \chandraacis\ data were obtained in 2000 (with ACIS-I3) and 2002 (with ACIS-S3) in faint mode. Our 2014-15 campaign (PI: Bogdanov) included 6 observations between September 2014 and February 2015 (Table~\ref{tab_xray_obs}). All observations in this campaign were performed in very faint mode on ACIS-S3. We focus our spectral analysis upon the 2015 \chandra\ observation; detailed analysis of other archival \chandra\ spectra is deferred to future papers.

\begin{figure}
\begin{center}
\includegraphics[scale=0.4]{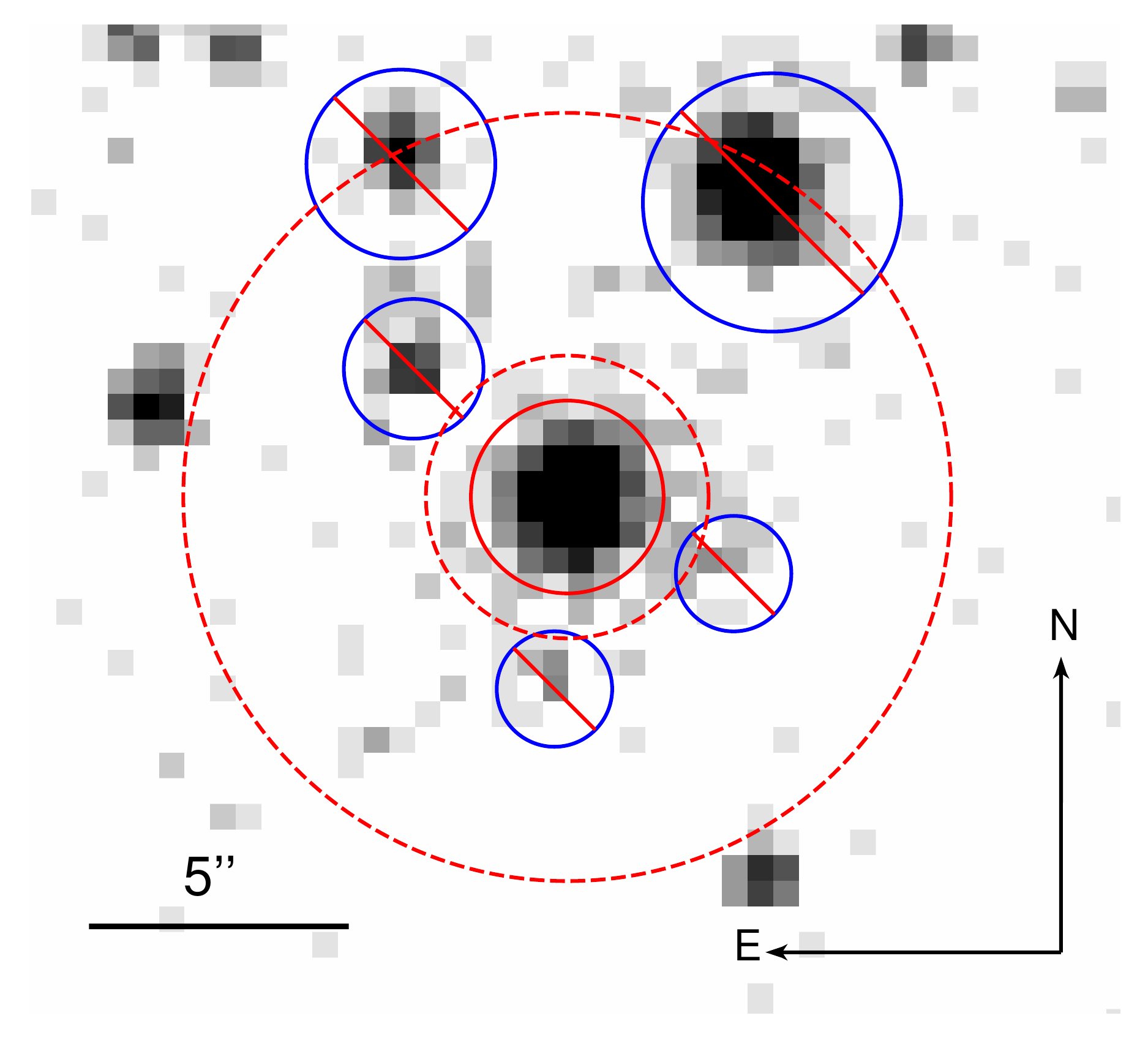}
\caption{\chandra\ 0.3-10 keV image of the vicinity of X9 in the core of 47 Tuc. Circles represent the nominal source (solid red circle) and background (dashed red circle) regions for spectral extractions of 47 Tuc X9. X-ray sources present in the background were excluded from the extraction region (blue circles).}
\label{fig_cxo_extraction}
\end{center}
\end{figure}

We used \texttt{CIAO} 4.7 with \texttt{CalDB} 4.6.8 \citep{Fruscione06} for data reprocessing and analysis. We reprocessed all \chandra\ data with \texttt{chandra\_repro} and extracted source and background spectra using \texttt{specextract}. We chose an extraction radius of 1.8$''$ for the source region.  For lightcurve analysis, we barycentered the time stamps.  47 Tuc X9 is  near the crowded center of the cluster. Thus, for extracting background spectra, we used an annulus with inner and outer radii of 2.7$''$ and 7.4$''$ around the source,  excluding detected sources (Fig.~\ref{fig_cxo_extraction}). Depending on exposure and data quality, we binned all final spectra to either 50 or 20 counts per bin. Finally we used XSPEC 12.8.2 \citep{Arnaud96} for spectral analysis, and performed spectral fitting in the 0.4--10 keV band.

\begin{table}
\centering
\begin{tabular}{@{}llll@{}}
\hline
Observatory/Inst.		&	Obs. ID		&	Date		&	Exposure (ks)\\
\hline
\hline
\Einsteinhri &	658		&	1979-04-21	&	1.6	\\
		&	4858		&	1979-11-18	&	0.8	\\
		&	4857		&	1979-11-19	&	1.5	\\
		&	4855		&	1979-11-19	&	2.0	\\
		&	4856		&	1979-11-19	&	1.9	\\
\hline
\Einsteinipc &	4969A	&	1979-11-19	&	4.8	\\
		&	4969B		&	1979-11-20	&	4.8	\\
		&	4969C		&	1979-11-21	&	7.2	\\
\hline
\rosathri &	300059N00	&	1992-04-19	&	1.2  \\
		&	300059N00	&	1992-05-21	&	3.4  \\
		&	300059A01	&	1993-04-17	&	13.2 \\
		&	400738N00	&	1994-11-30	&	18.9 \\ 
		&	400809N00	&	1995-10-25	&	4.6  \\
		&	400809A01	&	1996-11-16	&	17.5 \\
\hline
\chandraacis &	00953	&	2000-03-16	&	32	\\
		&	00955		&	2000-03-16	&	32	\\
		&	02735		&	2002-09-29	&	65	\\	
		&	02736		&	2002-09-30	&	65	\\
		&	02737		&	2002-10-02	&	65	\\
		&	02738		&	2002-10-11	&	69	\\
		&	16527		&	2014-09-05	&	41	\\
		&	15747		&	2014-09-09	&	50	\\
		&	16529		&	2014-09-21	&	25	\\		
		&	17420		&	2014-09-30	&	09	\\
		&	15748		&	2014-10-02	&	16	\\
		&	16528		&	2015-02-02	&	40	\\
\hline
\chandrahrc &	5542	&	2005-12-19	&	50	\\
		&	5543		&	2005-12-20	&	51	\\	 
		&	5544		&	2005-12-21	&	50	\\
		&	5545		&	2005-12-23	&	52	\\	 
		&	5546		&	2005-12-27	&	48	\\	 
		&	6230		&	2005-12-28	&	45	\\	 
		&	6231		&	2005-12-29	&	47	\\	 
		&	6232		&	2005-12-31	&	44	\\	 
		&	6233		&	2006-01-02	&	97	\\	 
		&	6235		&	2006-01-04	&	50	\\	 
		&	6236		&	2006-01-05	&	52	\\	 
		&	6237		&	2005-12-24	&	50	\\	 
		&	6238		&	2005-12-25	&	48	\\	 
		&	6239		&	2006-01-06	&	50	\\	 
		&	6240		&	2006-01-08	&	49	\\
\hline
\Swiftxrt &	84119001	&	2014-11-04	& 1.8	\\
		&	84119002	&	2014-11-17	& 4.2	\\
		&	84119003	&	2014-11-29	& 4.0	\\
		&	84119004	&	2014-12-11	& 4.2	\\
		&	84119005	&	2014-12-23	& 3.9	\\
		&	84119006	&	2015-01-07	& 3.7	\\
		&	84119007	&	2015-01-16	& 3.1	\\
		&	84119008	&	2015-01-28	& 3.3	\\
		&	84119009	&	2015-02-09	& 4.1	\\
		&	84119010	&	2015-03-05	& 3.8	\\
		&	84119011	&	2015-03-17	& 4.0	\\
		&	84119012	&	2015-03-29	& 3.9	\\
		&	84119013	&	2015-04-10	& 3.8	\\
		&	84119014	&	2015-04-22	& 3.6	\\
\hline
\nustar &	80001084002	&	2015-02-02	&	17	\\
		&	80001084004	&	2015-02-03	&	76	\\
\hline
\end{tabular}
\caption{The X-ray data used in this study. \Einsteinhri\ and \rosat\ data were obtained from \citet{Auriere89} and \citet{Verbunt98} respectively. All \chandra, \nustar\ and \Swiftxrt\ were reduced and analyzed in this study.}
\label{tab_xray_obs}
\end{table}

\subsection{NuSTAR}
\begin{figure*}
\begin{center}
\includegraphics[scale=0.8]{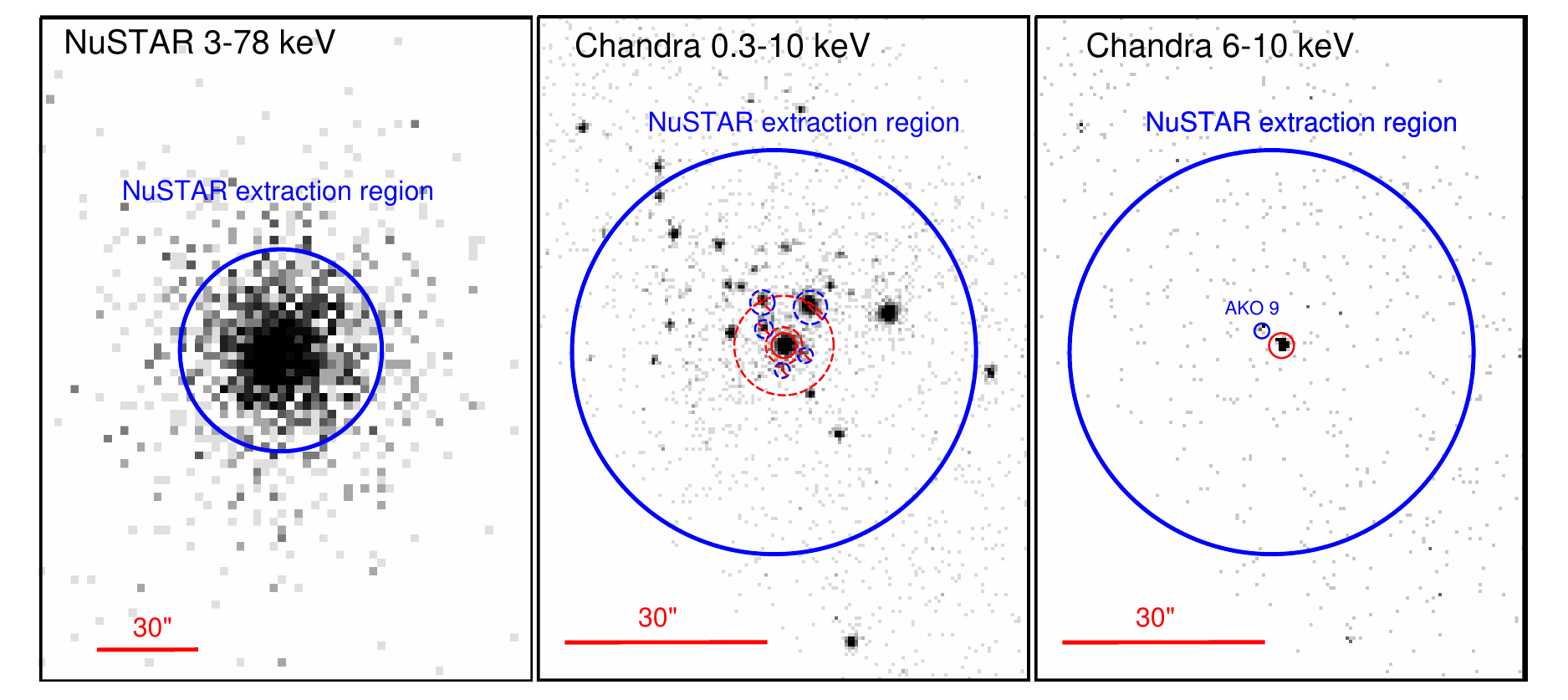}
\caption{X-ray image of X9 in 47 Tuc as observed by \nustar\ (left) and \chandraacis\ (center \& right). X9 is the only significant source above 6 keV in \chandra\ observations and thus is the major contributor in \nustar\ data of the cluster. The solid blue circle represents our spectral extraction region for X9 in \nustar\ data.}
\label{fig_cxo_and_nu}
\end{center}
\end{figure*}

\nustar\ has a relatively large point-spread function, with a half-power diameter of $\sim 1'$ \citep{Harrison13}, which encloses a substantial fraction of the core of 47 Tuc (Fig.~\ref{fig_cxo_and_nu}, left panel). 47 Tuc X9 is easily the brightest X-ray source above 2 keV  within the half-mass radius of 47 Tuc \citep{Heinke05a}. The CV AKO 9 \citep{Knigge03} is the only other source detected above 6 keV in our simultaneous \chandra\ observation. However, AKO 9 is $\sim$ 20 times fainter than X9 in the 6--10 keV band (Fig.~\ref{fig_cxo_and_nu}, middle and right panels). This makes X9 the principal source of X-rays above 6.0 keV in the cluster and ensures that \nustar\ spectra above 6 keV are minimally affected by confusion. We also checked the coordinates of the detected source, and find it consistent with the reported coordinates of X9.

Our \nustar\ observation of 47 Tuc was performed between February 2nd and 5th, 2015 and was simultaneous with \chandra\ ObsID 16528 and with our \ATCA\ radio observation.  Our \nustar\ observation was performed in two segments with a gap from Feb. 2nd, 21:11 to 3rd, 08:46 UT (Table~\ref{tab_xray_obs}).

X9 is clearly detected in our \nustar\ observations (Fig.~\ref{fig_cxo_and_nu}, left panel). The suggested extraction radius for weak sources in \nustar\ data is 30$''$(\href{https://heasarc.gsfc.nasa.gov/docs/nustar/nustar_obsguide.pdf}{\nustar\ observatory guide}). We used this radius for all our spectral extractions (source and background). We used the 6--79 keV band for all \nustar\ data analysis of X9.

We used the \texttt{nustardas} package released with \texttt{heasoft} 16.6 for data reprocessing and analysis. We processed the observations by running \texttt{nupipeline} and proceeded to extract spectra and light curves from both modules (FPMA \& FPMB) using \texttt{nuproducts}. After grouping the spectra by counts (20 per bin for the shorter segment and 40 for the longer segment), we used XSPEC for spectral analysis.

\subsection{Chandra/HRC}
The 47 Tuc \chandrahrc\ campaign (PI: Rutledge, \citealt{Cameron07}) contains 15$\times$50 ks observations performed over a period of 20 days in December 2005 and January 2006 (Table \ref{tab_xray_obs}). This provided us with a unique opportunity to study variations of X9 on time-scales of days. We combined these observations using \texttt{reproject\_obs} and extracted X9's light curve running \texttt{dmextract} (in CIAO software package) with background subtraction and assuming Gehrels uncertainties \citep{Gehrels86}. Finally, we barycentered the time stamps, and binned the light curve (if needed) into intervals of 10 ks, or 10 s (for different periodicity searches). 

\subsection{Swift/XRT, ROSAT/HRI, and Einstein}
As X9 shows strong variability in X-rays, we also investigated available archival data from \Swiftxrt, \rosathri, and \Einstein\ to study the extent of these variations.

Since 2014, \Swift\ has observed 47 Tuc frequently with long ($\sim 4000$ s) exposures as part of a GC monitoring programme\footnote{\url{http://www.iac.es/proyecto/SwiftGloClu/}} (Linares \& Chenevez, in prep.). Unlike typical \Swiftxrt\ exposures ($\sim$ 1000 - 2000 s), these exposures are adequate to constrain the brightness of sources at a few$\times 10^{33}$ erg/s at a distance of 4.5 kpc (like X9). Thus we analyze these datasets to look for variability and possible brightening (e.g., due to enhanced accretion) from X9. 

However, due to the relatively low resolution and large point-spread function of (FWHM $\sim7''$, \citealt{Burrows05}), the X-ray sources in 47 Tuc blend together in these observations. We chose to extract events in the 4--10 keV band, in which X9 is the dominant source, as a trade-off between signal/noise ratio and dominance of X9's flux over the rest of the cluster. After measuring background-subtracted count rates in the 4--10 keV band for each observation and subtracting contamination from other cluster sources in the \Swiftxrt\ extraction region (based on their flux in the 4--10 keV band \chandra\ image), we converted these count rates to flux by folding the best-fitting spectral model found for our 2015 \chandraacis\ + \nustar\ spectrum (see \S~\ref{sec_spec}) through the \Swiftxrt\ response matrix (using XSPEC) and finding equivalent flux values for each measured count rate. Ultimately we converted flux values to luminosity in the 0.5--10 keV band, assuming a distance of 4.53 kpc to 47 Tuc.

There are also archival \rosathri\ data available for 47 Tuc taken between 1991 and 1997 (0.5-2.5 keV), with spatial resolution ($\sim2''$) capable of marginally resolving the core of 47 Tuc (with radius of $\sim21.6''$). Detailed analyses of these data, attributing counts to each of the brightest core sources, have been reported by \citet{Verbunt98}. Thus we use the count rates reported by them \citep[][ figure 3]{Verbunt98} in our analysis. Similar to our procedure for \Swift\ data, we used the best-fitting spectral model folded to \rosathri\ to convert count rates to flux and luminosity. 

Among the instruments on board \Einstein, the High Resolution Imager (\emph{HRI}) was sensitive in the 0.15--3.0 keV band with a spatial resolution of $\sim2''$, and the Imaging Proportional Counter (\emph{IPC}) was sensitive in the 0.4--4.0 keV band with a resolution of $\sim1'$. Although \emph{HRI} had excellent spatial resolution, it had a low effective area (5 -- 20 cm$^2$). In contrast, \emph{IPC} had a relatively large effective area (100 cm$^2$). \Einstein\ observed 47 Tuc 8 times in 1979 (5 HRI observations followed with 3 IPC observations) and 1E 0021.8-7221 was clearly detected in all observations \citep{Hertz83,Auriere89}. We use the reported fluxes from \citet{Auriere89} in our analysis.

Although \rosathri\ and \Einstein\ have relatively poor angular resolutions, the high luminosity of X9 and its particularly hard spectrum compared to other sources in 47 Tuc's core allows us to use these observations to study X9.

All \Swiftxrt, \rosat\, and \Einstein\ observations used in this study are listed in Table~\ref{tab_xray_obs}.

\subsection{ATCA}
Since the existing radio and X-ray data on 47 Tuc X9 were taken several years apart \citep{Miller-Jones15}, we requested additional ATCA data to be simultaneous with the scheduled {\it Chandra} observations in 2015 February, under project code C3012.  The ATCA data were taken from 0505--1705 UT on 2015 February 2, with the array in an extended 6\,km configuration.  Using the Compact Array Broadband Backend \citep[CABB;][]{Wilson11} we observed simultaneously in two frequency bands, each of bandwidth 2048\,MHz, centred at 5.5 and 9.0\,GHz.  90\,min of data were lost to bad weather, leaving us a total of 523\,min on source.

We used B1934-638 both to set the amplitude scale and to determine the instrumental frequency response.  The starting elevation of the target was $52^{\circ}$, which decreased with time.  For the first 10\,hr of observations, we were able to use the same nearby compact source, B2353-686, to determine the time-variable complex gains as used by \citet{Miller-Jones15}.  However, that source set below the elevation limit of the telescope at 1445\,UT, after which we switched to the more distant, but higher-elevation complex gain calibrator J0047-7530 to enable continued observations of 47 Tuc.   We reduced the data according to standard procedures within the Multichannel Image Reconstruction, Image Analysis and Display (MIRIAD) software package \citep{Sault95}, and then exported the calibrated, frequency-averaged data to the Common Astronomy Software Application \citep[CASA;][]{McMullin07} for imaging. We co-added all the data and used a Briggs robust weighting of 1 to produce the image. We detected X9 in both frequency bands and used the CASA task IMFIT to fit a point source to the target in the image plane.  This gave measured flux densities of $27.6\pm7.2$ and $30.7\pm8.9$\,$\mu$Jy\,bm$^{-1}$ at 5.5 and 9.0\,GHz, respectively, giving a measured spectral index (defined such that flux density $S_{\nu}$ scales with frequency $\nu$ as $S_{\nu}\propto \nu^{\alpha}$) of $\alpha=0.2\pm0.8$. 

Comparing with observations in 2010 and 2013 \citep{Miller-Jones15},  the 9-GHz radio emission is similar, although the 5.5 GHz emission has dropped by a factor of 1.5$\pm$0.4.

\section{Results}\label{sec_results}
\subsection{X-ray variability}
\Einstein\ observations of 47 Tuc in the 1980s indicated that 1E 0021.8-7221 (the only detected X-ray source in the cluster at the time) showed strong variations. \citet{Auriere89} reported a variable absorbed flux of 2.2 - 6.4 $\times10^{-12}$ erg/cm$^{2}$/s in the 0.4-2.0 keV band. Assuming a distance of 4.53 kpc and extrapolating these luminosities to the 0.5-10 keV band using the power-law model reported by \citet{Auriere89}, with a photon index of 2, we find a luminosity range of 6.3 - 21.5 $\times10^{33}$ erg/s for \Einsteinhri\ and \Einsteinipc\ data points.

\begin{figure*}
\begin{center}
\includegraphics[scale=0.5]{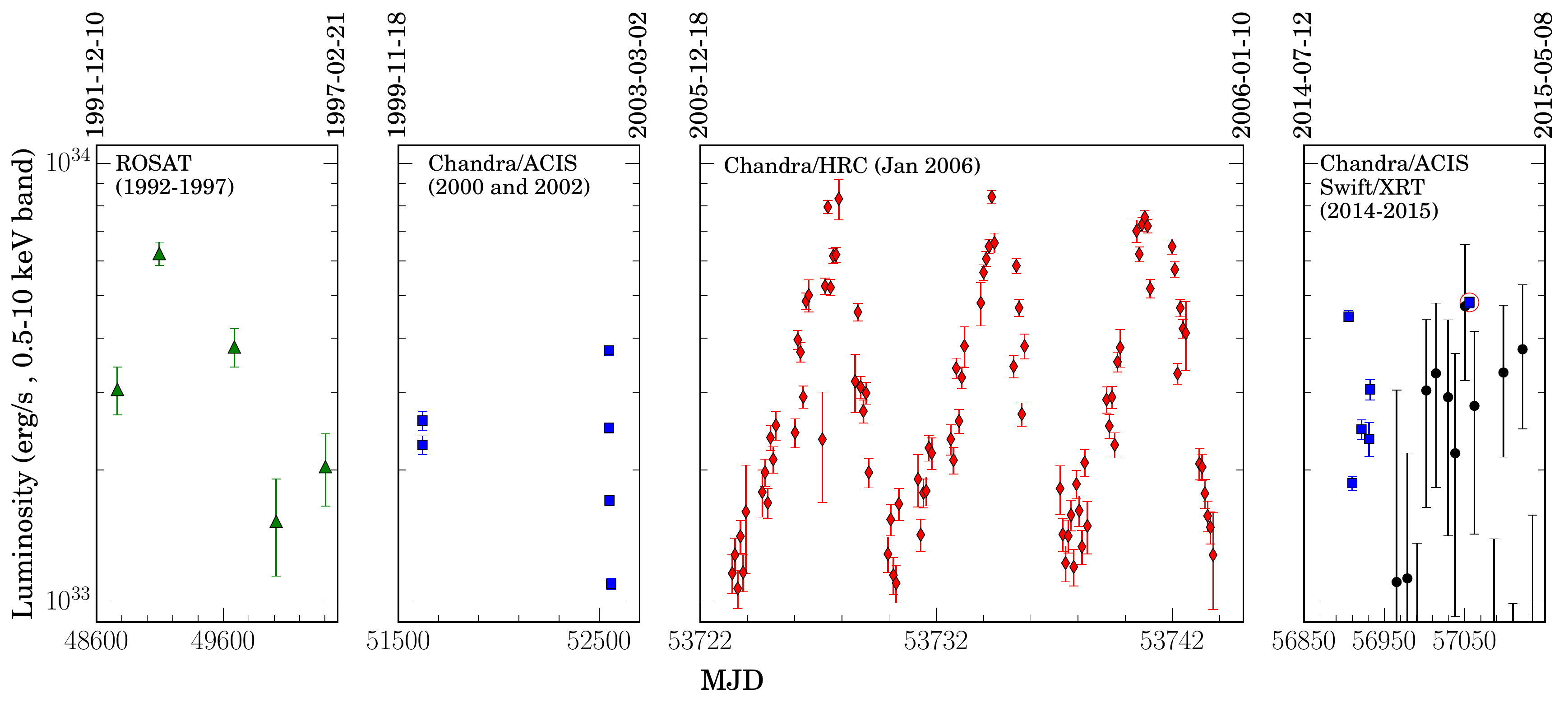}
\caption{Lightcurve of 47 Tuc X9 as seen by various telescopes since 1991. Note that each panel has a different timescale. Green triangles show \rosathri\ observations performed in the 1990s \citep{Verbunt98}, blue squares represent \chandraacis\ campaigns in 2000 \citep{Grindlay01a}, 2002 \citep{Heinke05a} and 2014-2015 \citep[][this work]{Bogdanov16}. Red diamonds show \chandrahrc\ observations between December 2005 and January 2006 \citep{Cameron07} and black circles show \Swiftxrt\ observations (Linares \& Chenevez in prep.). The \chandraacis\ observation performed simultaneously with \nustar\ and ATCA is indicated with a red circle around the blue square.}
\label{fig_lcs}
\end{center}
\end{figure*}

In \rosathri\ observations of 47 Tuc between 1992 and 1996 \citep{Verbunt98}, X9 showed a variable brightness between 1.5 and 6.1 $\times 10^{33}$ erg s$^{-1}$ (extrapolated to the 0.5-10 keV band). Similar variations can be seen in \chandraacis\ campaigns in 2000, 2002 and 2014-2015, and recent deep \Swiftxrt\ observations of the cluster (Fig.~\ref{fig_lcs}). 

\subsubsection{Super-orbital modulation}
\begin{figure*}
\includegraphics[scale=0.36]{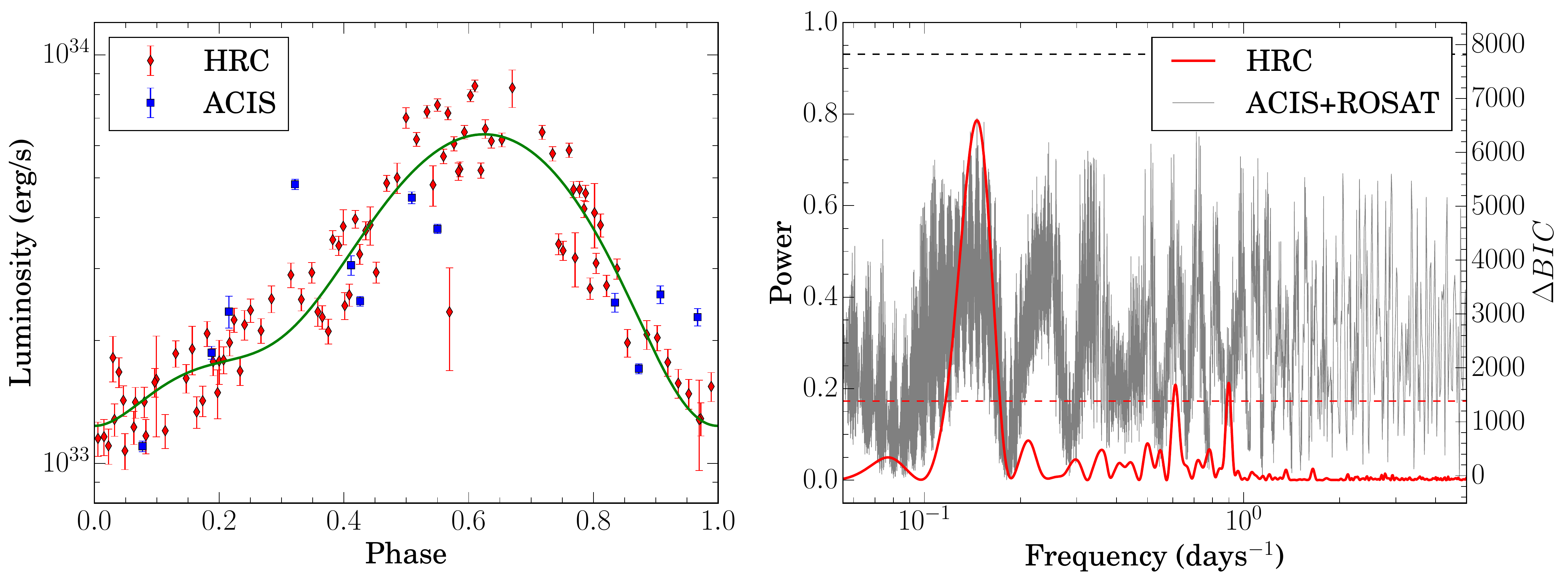}
\caption{{\it Left:} Folded X-ray lightcurve of X9 and a second order Fourier fit with a period of 6.8 days for \emph{HRC} and \emph{ACIS} data. The folded lightcurve of X9 shows non-sinusoidal behavior with a slower rise and a faster drop. 
{\it Right:} Result of a periodicity search using the Lomb-Scargle method for  \emph{HRC} data (red line) and for \emph{Chandra/ACIS} + \emph{ROSAT/HRI} data (gray line). The red and gray dashed horizontal lines indicate the 95\% confidence thresholds calculated using bootstrap resampling. Note that the uncertainties found with this method might be overestimated due to the non-sinusoidal profile of oscillations.}
\label{fig_LS}
\end{figure*}

The \chandrahrc\ lightcurve of X9 shows strong modulations between $1.1\times10^{33}$ erg s$^{-1}$ and $8.4\times10^{33}$ erg s$^{-1}$ on timescales of a few days (Fig.~\ref{fig_lcs}). We used a generalized Lomb-Scargle periodogram algorithm\footnote{Compared to the ``standard'' Lomb-Scargle periodogram, the ``generalized'' one utilized here allows for a constant offset term in the model and it also incorporates data uncertainties \citep[e.g., see][]{Zechmeister09,Vanderplas12} }. \citep{Lomb76,Scargle82} implemented in the Python package \texttt{AstroML}\footnote{\url{http://www.astroml.org/}} \citep{Vanderplas12,Ivezic14} to constrain the period of the variability observed in \chandrahrc\ data. We detected a formally significant periodic signal with a period of 6.8$_{-1.0}^{+1.8}$ days for \emph{HRC} data (Fig.~\ref{fig_LS}, right; red line).

Given the limited number of cycles observed in the light curve ($\sim$ three), we test if it is possible for the apparent periodicity to be a false detection, caused by red noise \citep[e.g.][]{Vaughan16}. Thus we check this possibility, by generating a large sample of red noise light curves following \citet{Timmer95}. We used the package \texttt{DELightcurveSimulation}\footnote{\url{https://github.com/samconnolly/DELightcurveSimulation}} \citep{Connolly15} to generate 50000 red noise light curves with a broken power-law power spectrum, with conservative power law indices of 1.1 and 2.2 \citep[e.g., see][]{Reig03}, placing the break at our detected period (as the most conservative case). We find that the probability of achieving a significant periodicity due to red noise with a power similar to (or higher than) that seen in our data is $0.07~\%$. Thus we conclude that the signal appears likely to be periodic, or quasi-periodic.

The uncertainties reported for the period here are calculated based on bootstrap resampling. A second order Fourier fit describes the light curve significantly better than a first order fit, with an F-test probability of chance improvement of $7\times10^{-11}$ \% for \emph{HRC} data\footnote{An F-test for the third order Fourier term provides a probability of 4.5\% for obtaining such improvement by chance, thus we prefer the two-term model.}. The two-term Fourier fit shows the rise towards the peak is slower than the fall (Fig.~\ref{fig_LS}, left).

The periodicity is only clearly visible in the \chandrahrc\ data. However, we also searched for indications of this variability in the \emph{ROSAT/HRI} and \emph{Chandra/ACIS} data. We performed a period search using the Lomb-Scargle method with \emph{ROSAT/HRI} and \chandraacis\ data points, excluding the \emph{Chandra/HRC} light curve. This search provides no clear periodic signal, but the maximum power in this periodogram is located at $P=6.8$ days (Fig.~\ref{fig_LS}, right). Although such a peak is insignificant by itself, given the consistency with the signal detected in \emph{Chandra/HRC} data, it adds suggestive evidence that the signal is real.

The statistical significance of this 6.8 day periodicity in the \emph{Chandra/HRC} data is strong enough that it is likely to be real, but this result is not conclusive.  The strong suggestion of a similar periodicity at different times in \emph{Chandra/ACIS} and \emph{ROSAT/HRI} data adds to the likelihood that the periodicity is real. We note the inherent problems in definitively identifying periods against red noise backgrounds with small numbers of cycles \citep{Press78,Vaughan16}, and we believe some caution is warranted in taking this signal as strong evidence of a periodicity. We consider the possible implications of such a periodicity, if confirmed, below (\S~\ref{sec_disc_suporb}). This period cannot be orbital, since the optical/IR counterpart is far too dim to allow a (subgiant) donor that would be needed to fill the Roche lobe with such an orbit \citep{Knigge08,Miller-Jones15}.

\subsubsection{Orbital periodicity}
\begin{figure*}
\includegraphics[scale=0.5]{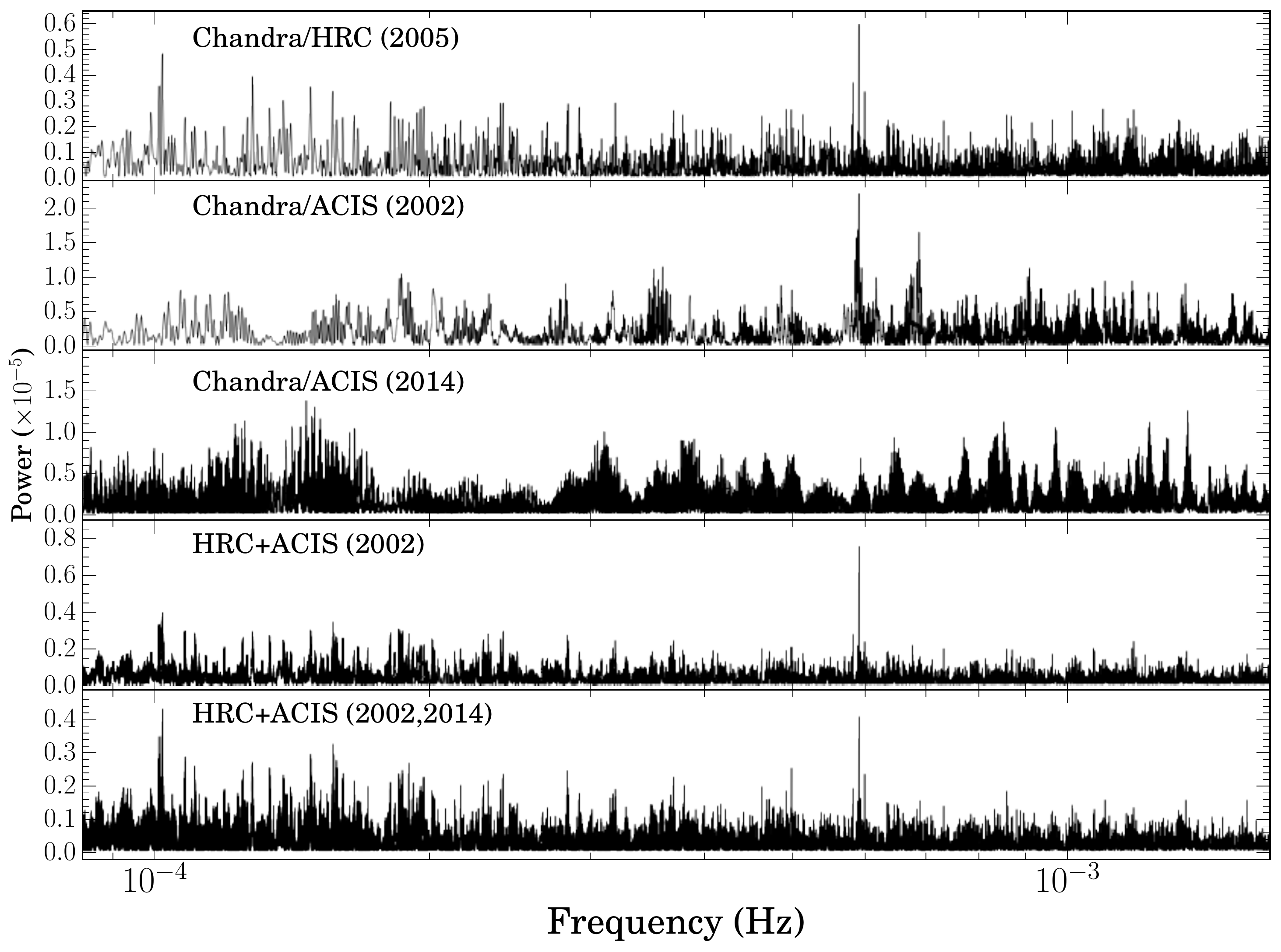}
\caption{\chandra\ power spectra for a periodicity search in the 10 min -- 4 hour range. A significant peak ($> 5\sigma$) is visible at $5.9\times10^{-4}$ Hz (28.2 min) in the \chandrahrc\ and \chandraacis\ 2002 light curves. Such a signal is not clearly detected in the \chandraacis\ 2014 light curve. The two bottom panels show the power spectra computed for combined light curves.}
\label{fig_orb_pwspec}
\end{figure*}

The long exposures of the \chandrahrc\ campaigns in 2005 enabled us to search effectively for signatures of periodic variability on short timescales (milliseconds - hours). We used a Lomb-Scargle periodogram to look for signs of an orbital period in the physically motivated range of 10 mins -- 4 hours (where the upper limit is set by the constraints on the donor from the UV/optical photometry, \citealt{Knigge08}). We detect a strong signal at 28.18$\pm0.02$ min (Fig. \ref{fig_orb_pwspec}, top panel). This motivated a search for similar signals in the \chandraacis\ light curves. We find a similarly strong signal in the 2002 \chandraacis\ 0.3 -- 10 keV light curve (Fig. \ref{fig_orb_pwspec}, second panel from top). However, this signal is not clearly detected in the 2014 \chandraacis\ light curve (Fig. \ref{fig_orb_pwspec}, middle panel). 

To ensure the detected signals in the \chandrahrc\ and \chandraacis\ (2002) light curves are not coincidentally matched noise peaks, we computed the power spectrum for the combined light curve. This significantly increased the significance of the signal (to higher than $5\sigma$), indicating that this signal is coherent, and real (Fig. \ref{fig_orb_pwspec}, second panel from the bottom). This detected periodic signal is in remarkable agreement with the predicted orbital period of $\sim$25 min by \citet{Miller-Jones15}, and supports the UCXB nature of this system. 

To determine the significance of the detected signal in the combined light curve, we simulated a large sample ($3.5\times10^6$) of white noise light curves based on our data and measured the peak amplitude in the power spectra. We neglect effects of red noise in these simulations, due to the large sampling baseline and the flat power spectrum in the physically motivated range (10 min - 4 hours). None of our simulated light curves show a peak amplitude similar to or higher than our actual data. Thus we infer a significance of higher than one in $3.5\times10^6$, or higher than $5\sigma$.

\begin{figure}
\includegraphics[scale=0.45]{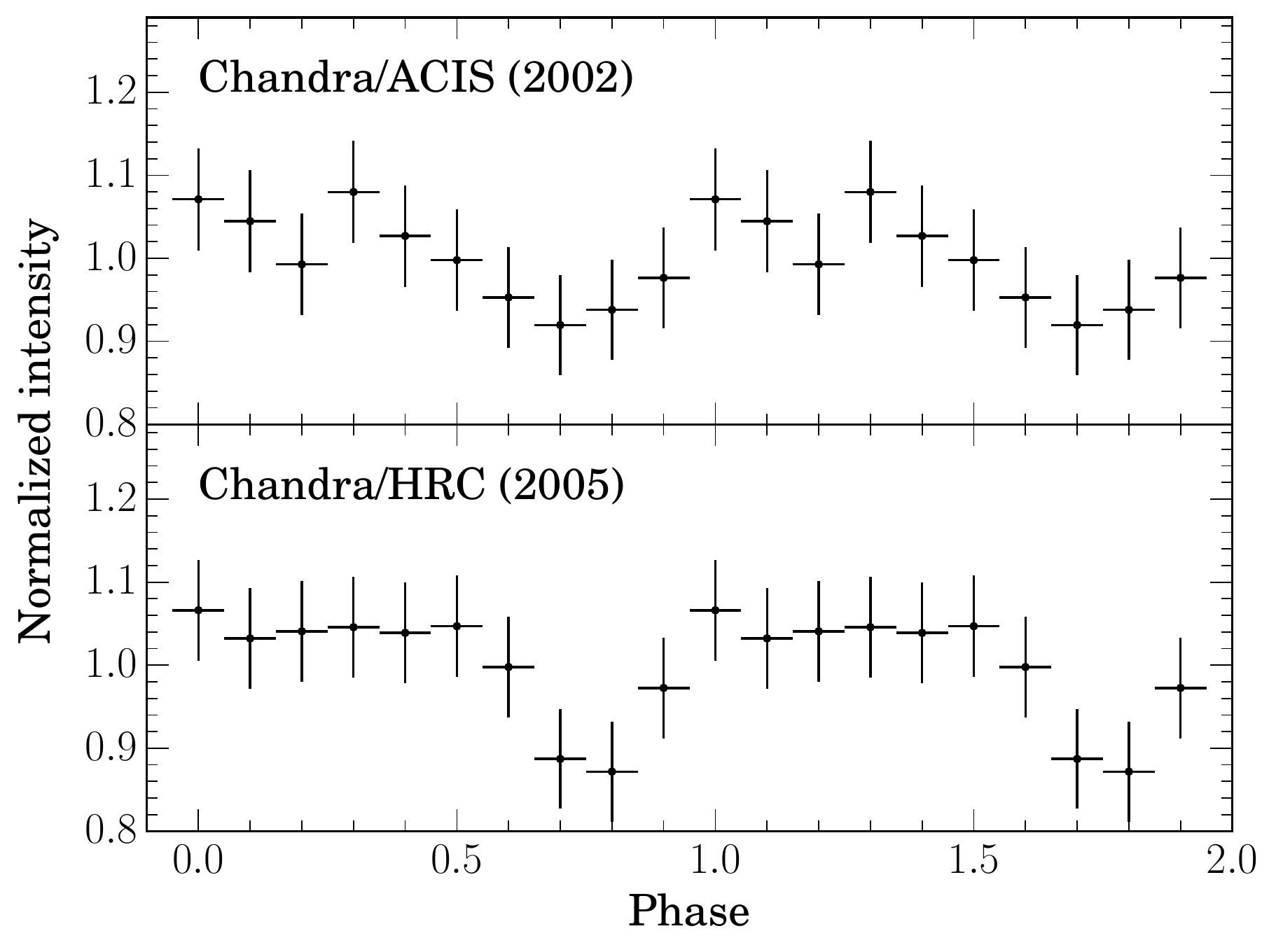}
\caption{\chandraacis\ (2002) and \chandrahrc\ light curves folded to a period of 28.2 min (1690.7738 s).}
\label{fig_orb_folded}
\end{figure}

The \chandrahrc\ and 2002 \chandraacis\ light curves, folded at the peak value of the period, are shown in Fig.~\ref{fig_orb_folded}. Both show evidence for a similar dip, lasting $\sim$30\% of the orbit, and reducing the flux by $\sim$15\%.

\begin{figure*}
\includegraphics[scale=0.441]{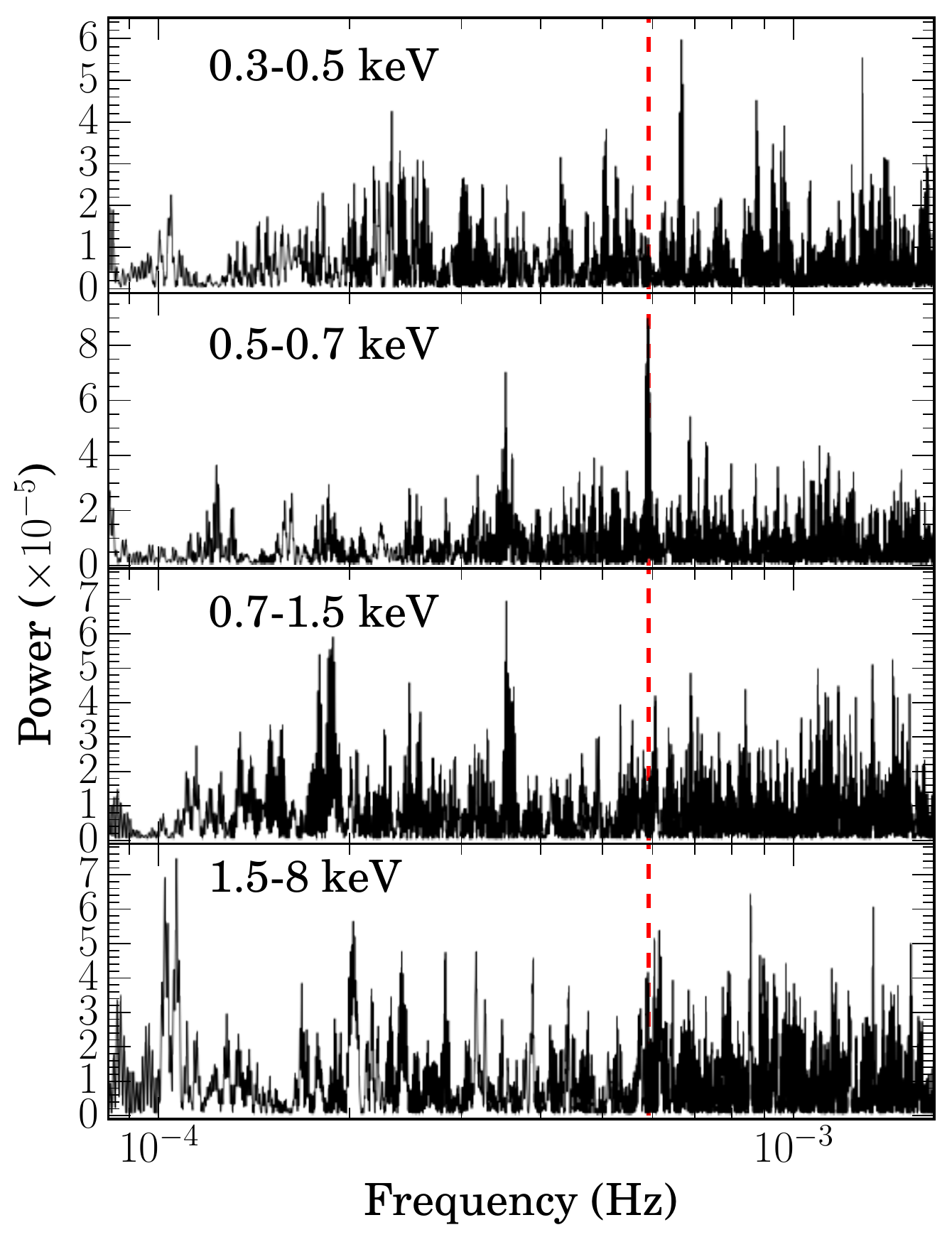}
\includegraphics[scale=0.55]{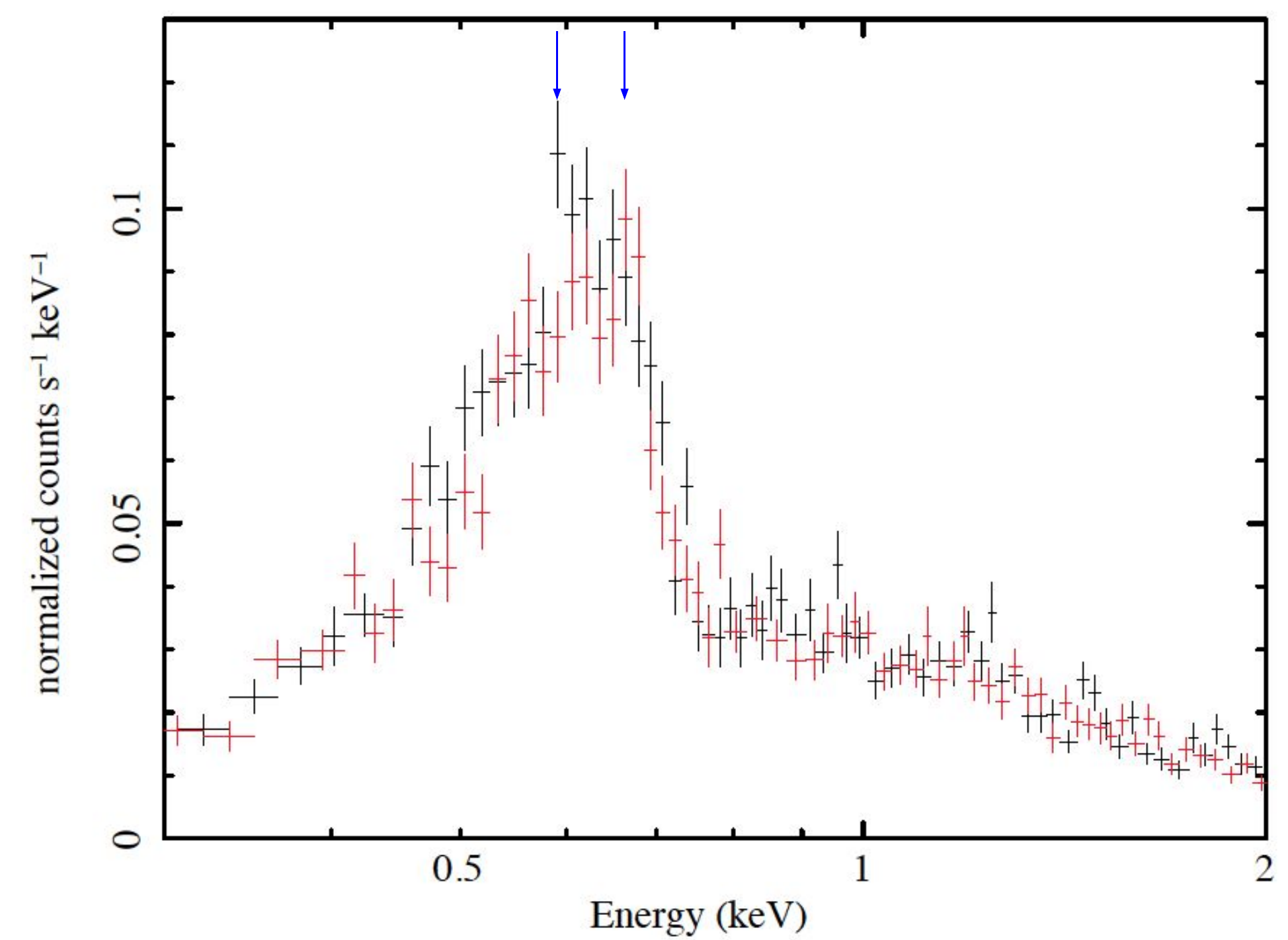}
\caption{\emph{Left:} Power spectrum computed from the \chandraacis\ 2002 light curve in different bands. The vertical red dashed line indicates the location of the detected peak in the full-band light curves. A peak is clearly visible in the 0.5 -- 0.7 keV band. However, there is a weaker signal in the 1.5 -- 8.0 keV band, indicating the signal is not exclusively present in one band. \emph{Right:} \chandraacis\ spectrum of X9 in the ``high'' and ``low'' phases. The chief difference appears in the oxygen lines (O VII at 0.57 keV and O VIII at 0.65 keV, indicated by blue arrows).}
\label{fig_orb_spec_powspec}
\end{figure*}

We computed power spectra for the \chandraacis\ 2002 light curve in different energy bands (0.3 -- 0.5, 0.5 -- 0.7, 0.7 -- 1.5, and 1.5 -- 7 keV) to determine the spectral component responsible for the periodic variations. We find 
that the periodic signal is strongly present (but not exclusively) in the 0.5 -- 0.7 keV band (Fig. \ref{fig_orb_spec_powspec}), where the dominant spectral features are two oxygen (VII and VIII) lines in the system. We discuss spectral analysis of the system and these lines in detail in \S \ref{sec_spec}. However, to identify the nature of the change in the spectrum causing the modulation, we extract spectra from the ``high'' and ``low'' phases as seen in the folded light curve (Fig.~\ref{fig_orb_folded}). We find that the most significant difference between the two phases is the relative strength of the oxygen lines. 

The absence of the periodic signal in the \chandraacis\ 2014/2015 campaign could be due to the reduced sensitivity of the \chandraacis\ detector \citep{Marshall04}. Between 2002 and 2014, the ACIS-S sensitivity below 1 keV dropped by a factor of $\sim3$ (See \chandra\ threads, ACIS QE Contamination\footnote{\url{http://cxc.harvard.edu/ciao/why/acisqecontam.html}}). Given that the amplitude of the variation is $\sim15\%$ and is dominantly below 1 keV, a drop by a factor of 3 in sensitivity is sufficient to weaken the signal. The shorter exposure of the 2014/2015 campaign (181 ks, compared to 264 ks in 2002 and 783 ks in 2005) could contribute as well.

\subsubsection{Search for other short-term variability}
We also searched for other short-timescale (milliseconds -- minutes) pulsations/modulations in the barycentred HRC data. Looking for second/minute periodic variability (e.g., the previously suggested 218 s modulation, \citealt{Grindlay01a}), we found no evidence of such modulations. 

Since X9 is in a compact binary, the rapid motion of a putative MSP in its orbit would spread the pulsed signal power over numerous Fourier bins in the power spectrum, which greatly diminishes its detectability. Moreover, due to the relatively low X-ray source count rate, it is not possible to search for millisecond pulsations over short segments of the orbit to minimize this effect. Therefore, to search for fast pulsations, we considered two Fourier-domain techniques from the PRESTO software suite \citep{Ransom02,Ransom03}. The "acceleration search" technique attempts to compensate for the large spin period derivative induced by the fast orbital motion of the pulsar.  This method is most effective when the duration of the observation is $\lesssim$25\% of the  orbital period.  The "sideband" (phase modulation) method, on the other hand, looks for sidebands produced by the binary motion in the power spectrum centered around the candidate pulsar spin frequency and coadds them in order to recover some sensitivity to the pulsed signal. This technique is most sensitive in situations where the observation is much longer than the orbital period and the observation is uninterrupted  \citep{Ransom03}. Given that the \textit{Chandra} HRC exposures are tens of kiloseconds, much longer than the apparent orbital period of X9, the phase-modulation method is more appropriate for this source.

For these searches it is necessary to bin the event data, which causes frequency dependent attenuation of the periodic  signal, with decreased  sensitivity  at  high  frequencies \citep[see, e.g.,][]{Middleditch76,Leahy83}. We searched each \textit{Chandra} HRC exposure separately, since the observations are too far apart to search them coherently for millisecond pulsations using these methods.  We found no statistically significant signal down to 1 ms (aside from the aliases of the 16 $\mu$s time sampling). To derive a pulsed fraction upper limit, we used tools available in PRESTO that consider the maximum power found in the power spectrum \citep[see][for details]{Vaughan94}. Based on the non-detection in the longest observation, we derive a $90 \%$ confidence upper limit of $19 \%$ on the pulsed fraction.

\subsubsection{Enhanced brightness in 1979}\label{sec_einstein}
One \rosathri\ observation of 47 Tuc in 1993 shows X9 at a luminosity of $6.1\pm0.4 \times10^{33}$ erg/s which, in comparison with later \rosathri\ and \chandraacis\ observations, was thought to be caused by an enhanced activity like a flare or outburst \citep{Verbunt98}. However, the \chandrahrc\ data shows that X9 reaches a peak luminosity of $\sim 7\pm0.2\times10^{33}$ erg/s periodically, and thus this level of brightness for X9 is not unusual.

However, the peak luminosities observed by \chandrahrc\ or \rosathri\ are $\sim$3.5 times fainter than the peak luminosity observed by \Einstein\ HRI observations in 1979 ($\sim 2.15\times10^{34}$ erg/s extrapolated to the 0.5 - 10 keV band). 
We note that due to the poor spatial resolution of \Einsteinipc\ (3 arcminutes), the IPC data analyzed by \citet{Auriere89} contains the entire half-mass radius of the cluster and may be the sum of flux from multiple sources. However, the peak of the lightcurve was recorded by the \Einsteinhri, which has superior spatial resolution ($\sim2''$), and localized the brightest \Einstein\ source to only $1.4''$ away from X9. 

\begin{figure}
\includegraphics[scale=0.53]{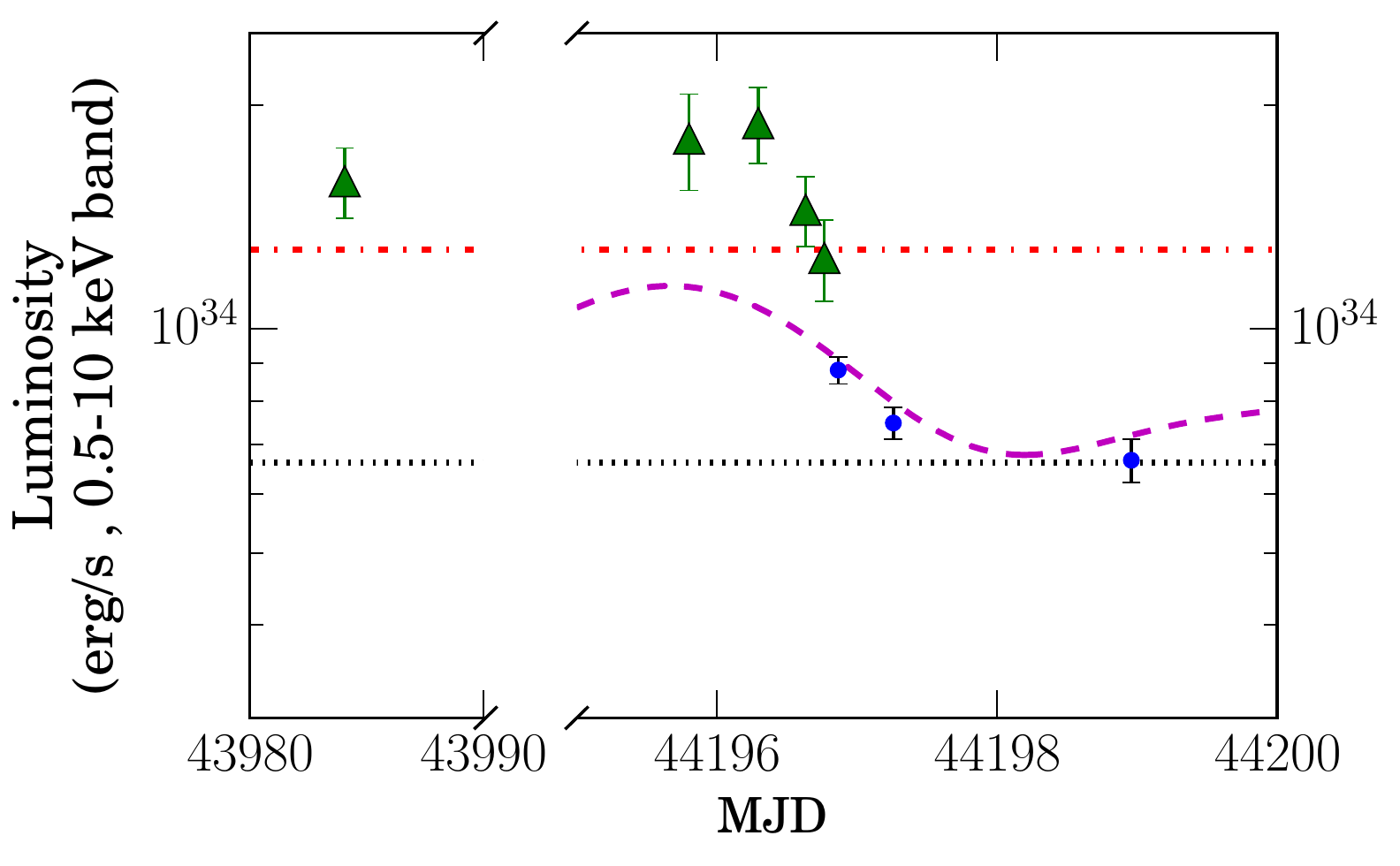}
\caption{\Einsteinhri\ (green triangles) and \Einsteinipc\ (blue dots) observations of 47 Tuc in 1979 as reported by \citet{Auriere89}. Black dotted line shows the cumulative luminosity of X-ray sources in 47 Tuc's half-mass radius with X9 at minimum luminosity, and the red dash-dotted line shows the cumulative luminosity of the cluster with X9 at the maximum luminosity observed by \chandrahrc. The dashed magenta line shows a projection of our Fourier fit to \chandrahrc\ data (with arbitrary shift applied in time). \emph{HRI} data points above the periodic profile and maximum observed luminosity indicate that an X-ray source in 47 Tuc was showing enhanced activity during \Einsteinhri\ observations.}
\label{fig_einstein}
\end{figure}

The peak of the combined \Einstein\ lightcurve, at $\sim2\times10^{34}$ erg/s, was also well above the typical cluster luminosity, $\sim 5.7\times10^{33}$ erg/s for all sources besides X9 \citep{Heinke05a}, and above the maximum cluster luminosity when X9 was at peak ($\sim1.3\times10^{34}$ erg/s). However, the \Einsteinipc\ data show a decline in luminosity down to fluxes consistent with the rest of the cluster, plus X9 in its lower state, on timescales typical of X9's variations (Fig.~\ref{fig_einstein}). The excellent match of the X-ray position, minimum flux, and decay timescale suggest that the bright state recorded by \Einsteinhri\ was due to X9, reaching a peak luminosity roughly twice the maximum luminosity seen by \rosat, \chandra\ or \Swift\ later.

\subsection{X-ray spectroscopy}\label{sec_spec}
We tried to constrain the broadband spectral shape through joint \chandra\ and \nustar\ spectral fitting. We used XSPEC 12.8.2 in all our spectral analysis and assumed \citet{Wilms00} abundances and \citet{Verner96} cross-sections. Foreground reddening in the direction of 47 Tuc is $E(B-V)=0.04$ \citep{Salaris07}. Assuming $R_V = 3.1$ and using the observed correlation between extinction ($A_V$) and hydrogen column density $N_H$ for Wilms abundances \citep{Bahramian15,Foight16}, we estimate a hydrogen column density of $3.5(\pm0.1)\times10^{20}$ cm$^{-2}$. We assumed this as the hydrogen column density towards 47 Tuc throughout this paper. Since $N_H$ is quite small, variations in hydrogen column density (e.g., due to differential reddening) would lead to correspondingly small variations in spectral fits. For all spectral analyses, we considered 0.4-10 keV band data from \chandraacis\ and 6.0-79 keV data for \nustar. Our \nustar\ spectra contain data from two observations (Table \ref{tab_xray_obs}) and two modules (FPMA \& FPMB), We fit all these spectra jointly (instead of combining the spectra). Due to the brightness of X9, we also accounted for the effects of pileup on \emph{ACIS} data by including the \texttt{pileup} model in XSPEC in all \chandra\ spectral fittings, even though the 2014-15 \chandraacis\ campaign used a 1/8 sub-array mode to reduce pileup (to less than $2 \%$). Due to possible differences in detector responses between \nustar\ and \chandraacis, we include a variable coefficient in all our models representing the \nustar/\chandra\ relative normalization ratio.

The X-ray spectrum of X9 (Fig.~\ref{fig_spec}) contains a strong line complex in 0.5--0.7 keV, attributed to O VII (at 0.57 keV) and O VIII (at 0.65 keV) by \citet{Heinke05a}, and an extended hard energy tail. We see no evidence of a K$\alpha$ iron line. Our initial model for fitting the spectrum was an absorbed collisionally-ionized gas model \citep[\texttt{mekal} model in XSPEC, e.g.,][]{Mewe85,Mewe86}, plus a cut-off power law (\texttt{constant*pileup*tbabs*(mekal+cutoffpl)})\footnote{Note that the constant factor in the model represents the \nustar/\chandra\ normalization ratio.}.  However, this model failed to describe the spectra ($\chi_\nu^2 = 1.56$ for 185 d.o.f). In particular, it failed to fit some of the stronger oxygen features below 1 keV. We discuss fit improvements for these features in detail in Sections \ref{sec_mekal}, \ref{sec_refl} \& \ref{sec_phoion}.

\subsubsection{Comptonization and reflection}\label{sec_comp}
In addition to the low energy features, our initial fit (\texttt{constant*pileup*tbabs*(mekal+cutoffpl)}) shows a significant trend of residuals as a ``bump'' in the 10-30 keV region (Fig.~\ref{fig_spec}, top-left). Thus we test other appropriate models by replacing the power law component with them. To compare models in how they fit the harder part of the spectrum, in this section we report statistics in the 1-79 keV band for each model. Our initial power-law-only model yielded a $\chi_\nu^2$ of 1.37 for 171 d.o.f. Replacing the power-law with a Comptonization model \citep[e.g., \texttt{compTT};][]{Titarchuk94} does not improve the fit ($\chi_\nu^2$ of 1.39 for 170 d.o.f). 

It is possible that the observed bump is caused by reflected emission. Replacing the power law with a reflected power law from neutral matter \citep[\texttt{pexrav};][]{Magdziarz95} improved the fit significantly ($\chi_\nu^2 = 0.98$ for 168 d.o.f, Fig.~\ref{fig_spec}, top-right). A reflected power-law from ionized matter (\texttt{pexriv}) gives similar results ($\chi_\nu^2= 0.99$ for 166 d.o.f) but leaves the disk temperature and ionization parameter ($\xi = 4 \pi F_{ion} /n$, where $F_{ion}$ is irradiating flux in the 5-20 keV band and $n$ is density of the reflector) mostly unconstrained ($\xi<270$ erg cm/s). We also tried fitting the hard component of the spectrum using a Compton-scattering model convolved with a reflection model (e.g., \texttt{reflect*comptt}). This model also addresses the reflection bump well ($\chi_\nu^2= 1.00$ for 167 d.o.f). The details of our spectral fitting with these models are tabulated in Table~\ref{tab_spec_hard}. 

The general improvement achieved by adding effects of reflection to the model, and the fact that the reflection scale parameter in all cases is non-zero ($>0.5$), give strong evidence for the significance of reflection in this system.

All three models containing reflection fit the spectrum well. We prefer a non-ionized reflecting material, as the ionization parameter in Pexriv is loosely constrained and is consistent with zero (i.e., suggesting insignificance of ionized matter in the reflection). As the physically motivated model (reflect$\times$compTT) requires an unusually high reflection scale parameter, indicating significant obscuration of the emission source (which seems inconsistent with the constraints on the inclination angle in the reflection component), we use Pexrav for the rest of our analysis. 

\begin{table*}
%\centering
\begin{tabular}{lllllll}
\hline
				&\multicolumn{2}{c}{Without reflection} &	& \multicolumn{3}{c}{With reflection}\\
\cline{2-3}   \cline{5-7}  
				&	Powerlaw	&	CompTT				&	&	Pexrav				&	Pexriv				&	Reflect$\times$CompTT	\\
\hline
photon index	& 1.01$\pm0.07$ &	--					&	& 1.2$\pm0.1$			& 1.17$_{-0.03}^{+0.1}$ &	--						\\	
E$_{fold}$ (keV)& $>85$			& 	--					&	& 56$_{-14}^{+22}$		& 62$_{-15}^{+63}$		&	--						\\
T$_0$ (keV)		& --			& $<0.3$				&	&	--					&	--					& 0.35$\pm0.04$				\\
T$_e$ (keV)		& --			& $>161$				&	&	--					&	--					& 62$_{-31}^{+234}$			\\
$\tau$			& --			& 1.8$_{-0.8}^{+1.3}$	&	&   --					&	--					& 0.9$_{-0.8}^{+2.7}$		\\
Reflect scale	& --			&	--					&	& 1.3$_{-0.4}^{+2.0}$	& 1.1$_{-0.3}^{+1.0}$	& 5.5$_{-1.8}^{+4.4}$		\\
Fe. abund.		& --			&	--					&	&	$<0.3$				&	$<0.3$				& $<0.3$					\\
$\theta^{\circ}$& --			&	--					&	&	$<68$				&	$<45$				& $<60$						\\
T$_{disk}$ (keV)& --			&	--					&	&	--					&	0.001$\pm?~^a$		&	--						\\
$\xi$ (erg cm/s)& --			&	--					&	&	--					&	$<270$ 				&	--						\\
C$_{NU/CX}$		& 1.03$\pm0.09$	& 1.07$_{-0.07}^{+0.09}$&	& 0.90$\pm0.09$			& 0.88$_{-0.09}^{+0.1}$	& 0.87$_{-0.06}^{+0.13}$	\\
$\chi_\nu^2$/d.o.f&	1.37/171  	& 1.39/170				&	& 0.98/168				& 0.99/166				& 1.00/167					\\
N.H.P (\%) 		& 0.06			&	0.05				&	&	54.2				&	52.4				&	47.2					\\
\hline
\end{tabular}
\caption{Best-fit parameter values for hard spectral components, fit in the 1-79 keV band. T$_0$ is the temperature of the incident photons, T$_e$ is the electron temperature and $\tau$ is optical depth in Comptonization models. We assume a disk geometry for the Comptonization component (CompTT). E$_{fold}$ is the folding energy for exponential cut-off in the power law tail. Reflection scale indicates the relative scale of the reflected emission compared to the source. Iron abundance is indicated as a fraction of solar abundance. The inclination angle is denoted by $\theta$ and $\xi$ represents the ionization parameter. C$_{NU/CX}$ shows the relative \nustar\ to \chandraacis\ normalization factor. N.H.P. indicates the null hypothesis probability. $^a$- The disk temperature in the Pexriv model is not constrained and the value stated here is just from the best fit.}
\label{tab_spec_hard}
\end{table*}

\subsubsection{Collisionally-ionized gas models}\label{sec_mekal}
After addressing the hard component of the spectrum, we focused on the features in the soft end while fitting the whole band (0.4-79.0 keV). Our model with a single Mekal model and pexrav (\texttt{constant*pileup*tbabs*(mekal+pexrav)}) failed to describe the strong emission features below 1.0 keV ($\chi_\nu^2 = 1.34$ for 183 d.o.f) leaving significant residuals below 1 keV. Adding a second mekal component did not improve the fit. As the features we are trying to fit are dependent on the abundances of elements in the system, we test our model with 47 Tuc abundances of elements (following \citealt{Heinke05a}, assuming linear abundance of 60\% solar for C, N, and O, 40\% solar for elements Ne through Ca, and 20\% solar for Fe and Ni, see also \citealt{Carney96,Salaris98}). To do so, we replace the Mekal component in our model with VMekal, which allows for adjustable abundances of elements. This model provides a slightly better fit with a $\chi_\nu^2 = 1.23$ for 183 d.o.f (Fig.~\ref{fig_spec}, middle-left). Adding a second vmekal component did not improve the fit.

The major features that the models above fail to describe mainly consist of strong spectral lines, from O VII and VIII between 0.6 and 0.8 keV (Fig.~\ref{fig_spec_model}). Additionally, in our \chandraacis\ spectrum, there is a bump at $\sim0.3$ keV which is not fit well when our band is extended to 0.3 keV. This bump could be due to uncertainties in the \emph{ACIS} response matrix, or could be due to carbon lines. This is particularly plausible as \citet{Knigge08} identify strong carbon lines in the FUV spectrum of X9. This significant underestimation of oxygen emission in our models suggests that the abundance of oxygen (and possibly carbon) in this system is higher than both the measured values for 47 Tuc, and solar abundances. \citet{Miller-Jones15} suggest that X9 might be a UCXB (based on mass transfer rate). This is consistent with possibly high abundance of oxygen, as in UCXBs where the donor is a C/O WD, carbon and oxygen can be dominant species. We attempt a first test of this scenario by using the abundances calculated by \citet{Koliopanos13} for a C/O UCXB system.  Although a model with a single VMekal component with UCXB abundances fails to describe the features below 1 keV ($\chi_\nu^2 = 1.35$ for 183 d.o.f), adding a second VMekal component gives a slightly better fit with $\chi_\nu^2 = 1.22$ for 181 d.o.f. (Table~\ref{tab_spec_soft} and Fig.~\ref{fig_spec}, middle-right). The slight improvement in the fit and the less-structured appearance of the residuals suggests a C/O WD companion in this system.

\subsubsection{Reflection models for UCXBs}\label{sec_refl}
Our spectral analysis in Section \ref{sec_comp} provides evidence for reflection, and Section \ref{sec_mekal} indicates an ultra-compact nature for the system, with a high abundance of oxygen. Thus it is possible that the strength of oxygen lines is due to reflection from a hot, oxygen-rich disk. We test this scenario by fitting the spectrum with a reflection model appropriate for oxygen-rich UCXBs\footnote{None of the standard reflection models incorporated within XSPEC consider hydrogen-deficient gas.}. For this purpose we used Xillver$_{\text{CO}}$ \citep{Madej14}, which is a reflection model for accretion disks in oxygen-rich UCXBs. Xillver$_{\text{CO}}$ is based on the Xillver reflection model \citep{Garcia10,Garcia11,Garcia13}.

Xillver$_{\text{CO}}$ contains only the \emph{reflected} emission from a disk plus a power law continuum (and not the emission from these components themselves). Thus to include all the components, our model became \texttt{constant*pileup*tbabs*(diskbb+cutoffpl+xillverCO)}. In this model we tied the disk temperature between Xillver$_{\text{CO}}$ and diskbb, and tied the photon index and cut-off energy between Xillver$_{\text{CO}}$ and the cutoff powerlaw. This model gives a moderate fit with $\chi_\nu^2 = 1.30$ for 181 d.o.f. However it shows clear residuals in fitting the oxygen features and the reflection bump (Fig.~\ref{fig_spec}, bottom-left).

\subsubsection{Photo-ionized gas models}\label{sec_phoion}
None of the models tested in Sections ~\ref{sec_mekal} \& \ref{sec_refl} seem to be able to entirely address the oxygen features. Specifically, most of these models seem to underestimate the relative amplitude of O VIII compared to O VII (these lines are demonstrated in Fig.~\ref{fig_spec_model}). This could be due to significant increase in the ionization state of oxygen due  to photo-ionization. To investigate this, we used XSTAR \citep{Kallman01} to simulate a model spectrum of photo-ionized oxygen in a system with a size scale applicable to X9. XSTAR is designed to self-consistently simulate emission and absorption by hot, diffuse gas. We assume that the diffuse gas surrounds the system and is illuminated by the inner accretion flow.

For our simulations we considered two scenarios: (i) a diffuse gas region filled with just carbon and oxygen, and (ii) a purely oxygen diffuse gas region. In both cases we assumed the gas is illuminated by a source (possibly the inner accretion flow) with an X-ray luminosity of $\sim 10^{33}$ erg/s (a typical luminosity level for X9). For the illuminating source we assumed a power law spectrum obtained from the best power law fit to the \chandraacis\ spectrum in the 1-10 keV band (with a photon index of 1.1 and a 0.5-10 keV flux of 1.4$\times10^{-12}$ erg/s/cm$^2$). For the simulations, we assumed a hydrogen column density of 10$^{22}$ cm$^{-2}$ and a constant gas density of 10$^{10}$ cm$^{-3}$. While quick tests indicate our results do not appear to be highly sensitive to these parameters, we will investigate a broader range for these parameters in future work. The variables in our grid model were temperature (10$^4$ to 10$^7$ K), ionization parameter\footnote{Note that this ionization parameter is defined as $\xi = L/nR^2$ where L is luminosity, n is density and R is distance from the illuminating source.} ($\log \xi$ from -4 to 4) and in case of the C/O gas, relative abundance of carbon compared to oxygen (from 0 to 2). Our simulations were parallelized using PVM\_XSTAR \citep{Noble09}. Finally we produced XSPEC table models using the XSTAR task \texttt{xstar2table} and performed spectral fitting with these models. From here on, we refer to these models as the photo-ionization model (\texttt{phoionize}).

We used a slightly different version of XSTAR (2.2.1bo8) compared to the regular XSTAR release (2.2.1bn24). The main difference between the two versions is the effects of bound-bound radiative excitation which is turned on in the 2.2.1bo8 release. This is especially important as the oxygen Ly $\alpha$ line can also be emitted by bound-bound radiative excitation of the upper level by continuum photons. Thus the 2.2.1bo8 release produces the helium-like oxygen line at a different energy as it pumps the resonance line, while the regular version makes a stronger intercombination line via recombination.
 
\begin{figure*}
\begin{center}
\includegraphics[scale=0.33]{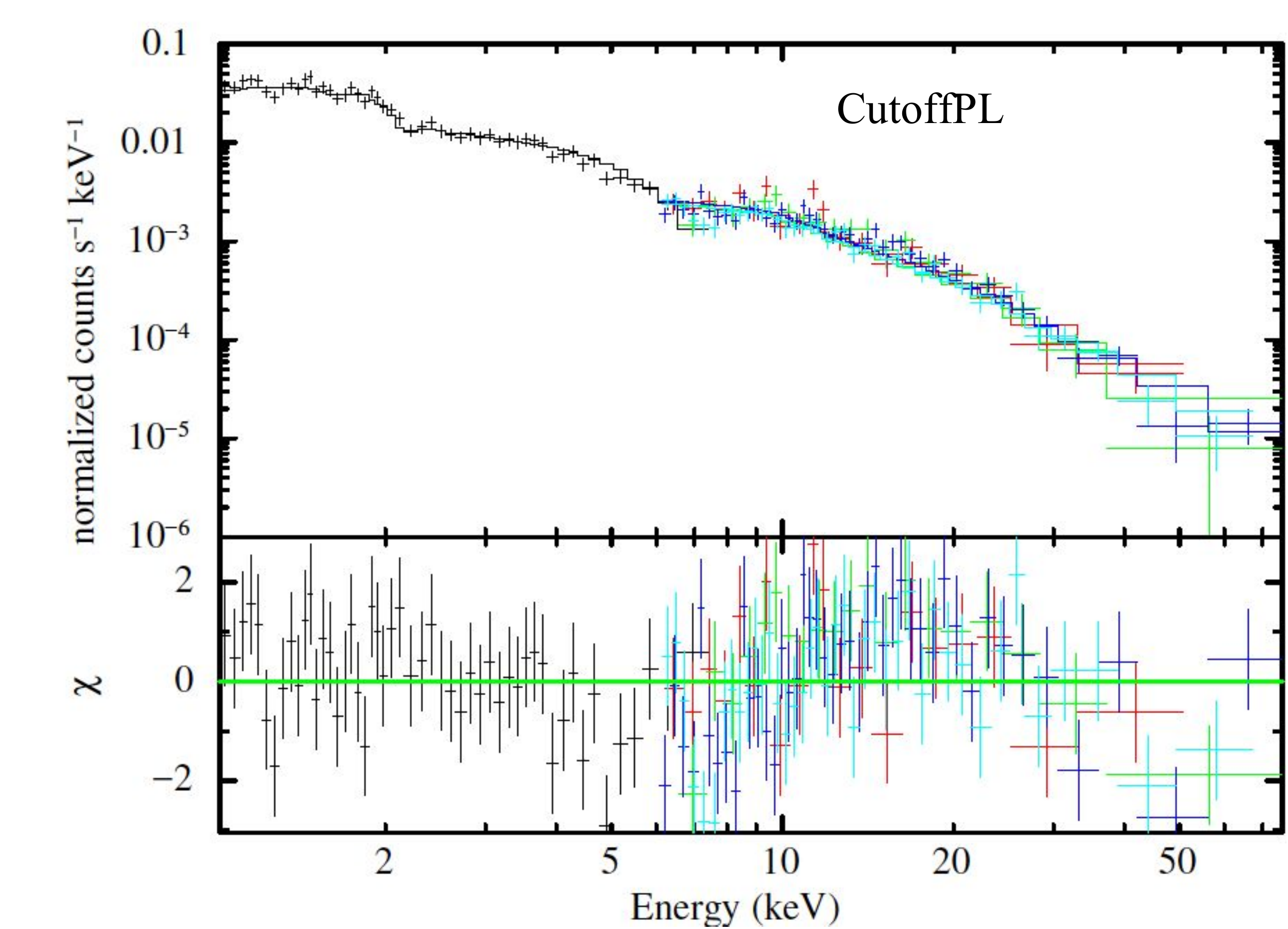}
\includegraphics[scale=0.33]{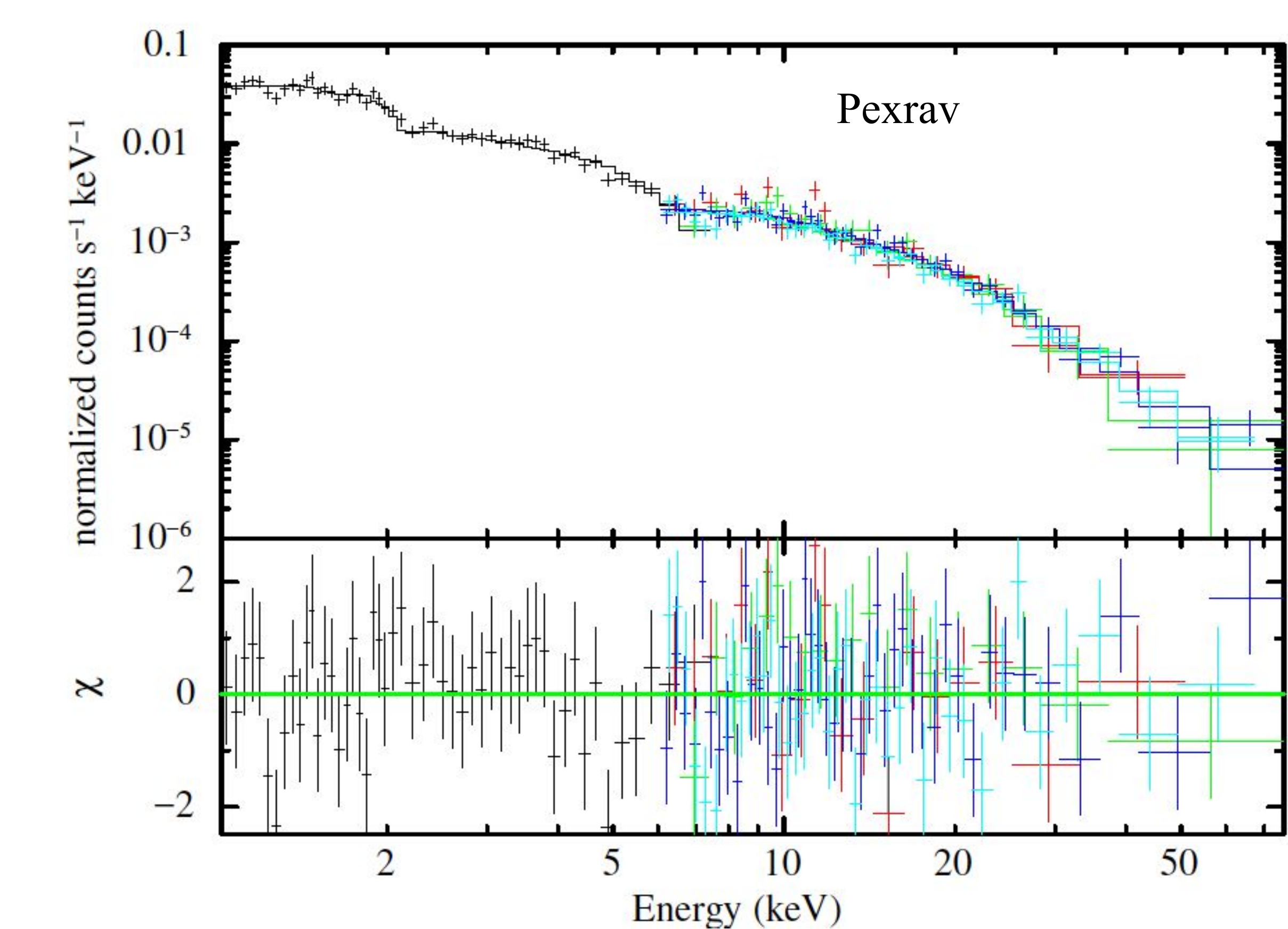}
\includegraphics[scale=0.33]{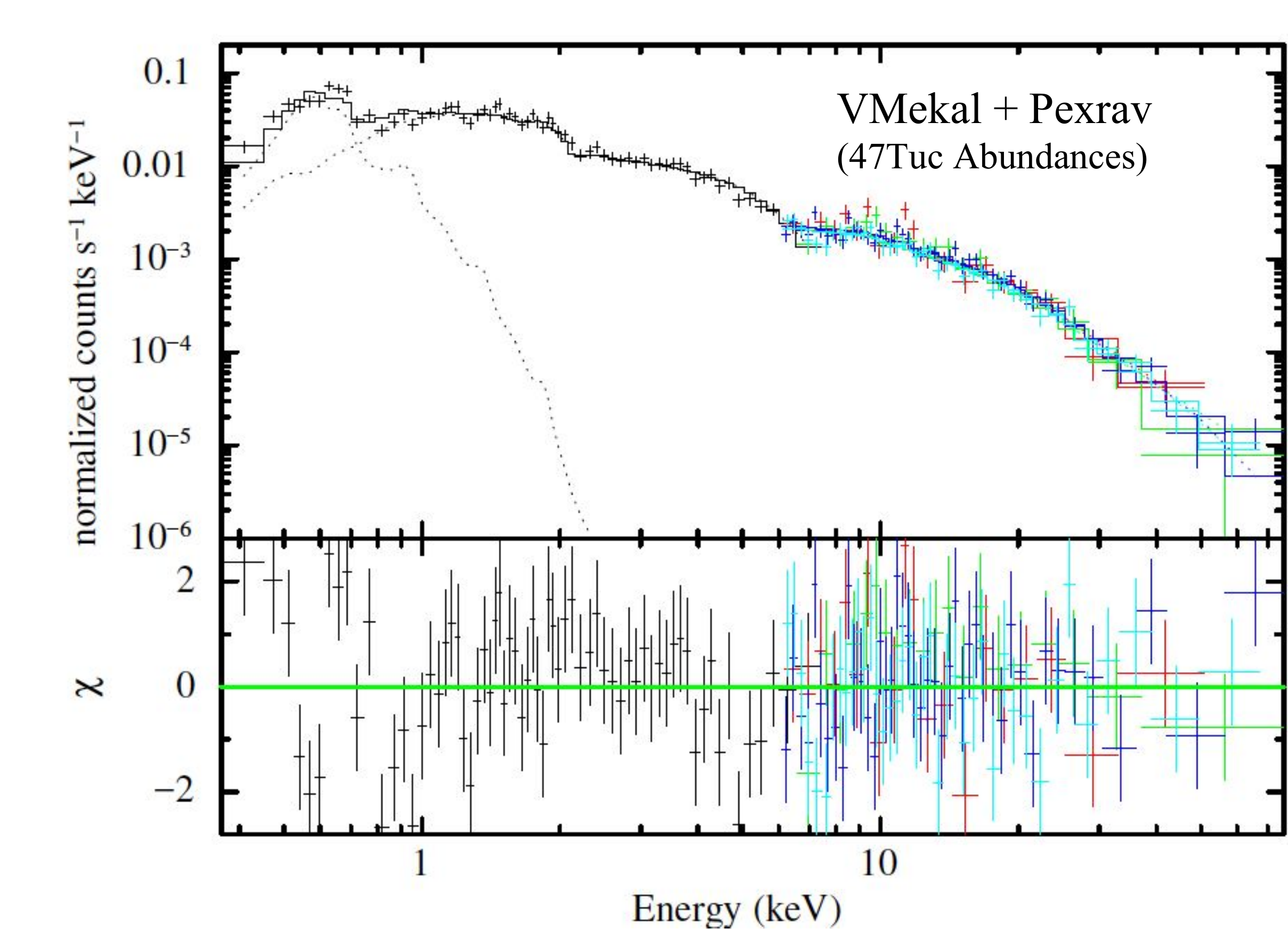}
\includegraphics[scale=0.33]{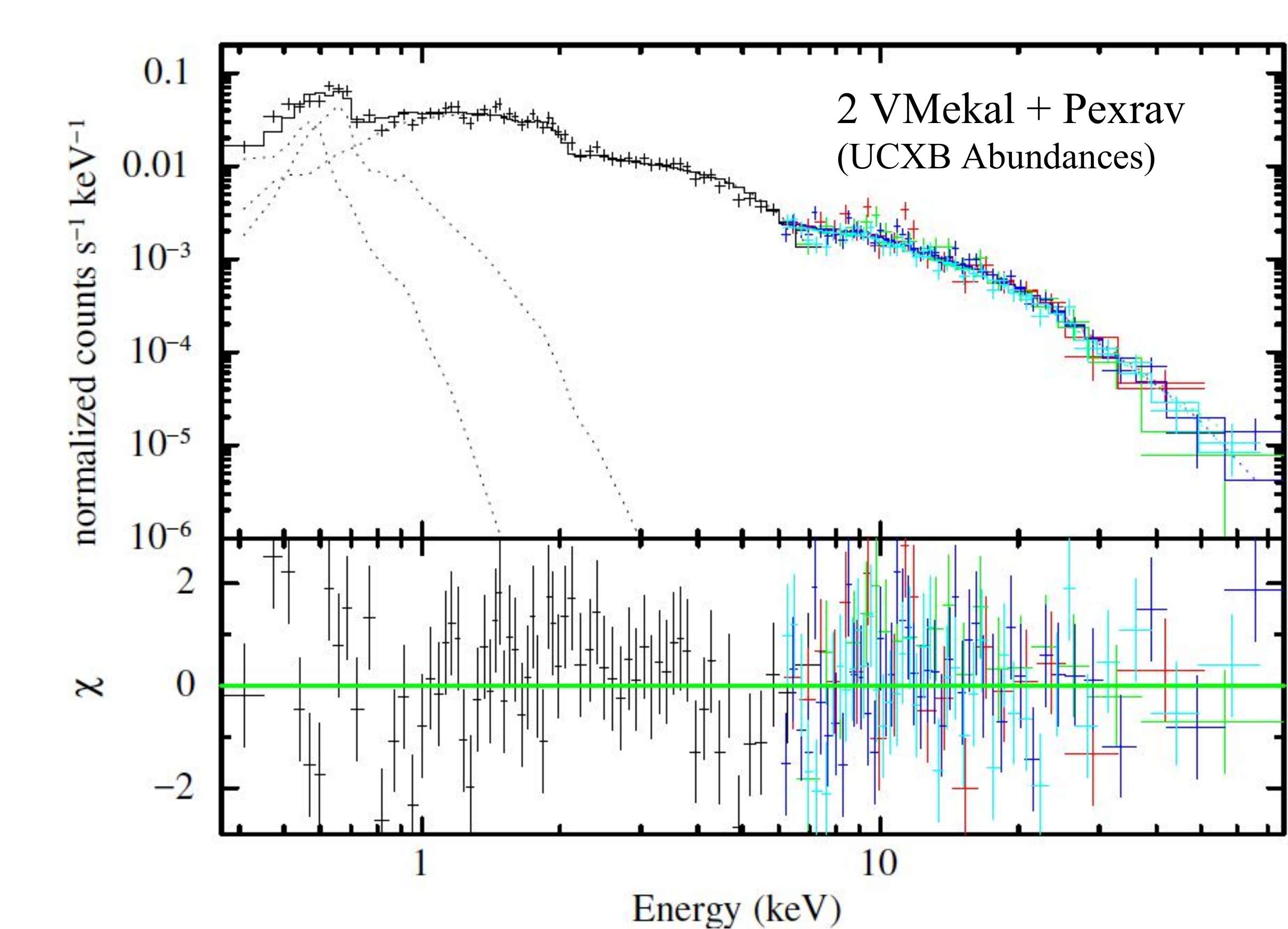}
\includegraphics[scale=0.33]{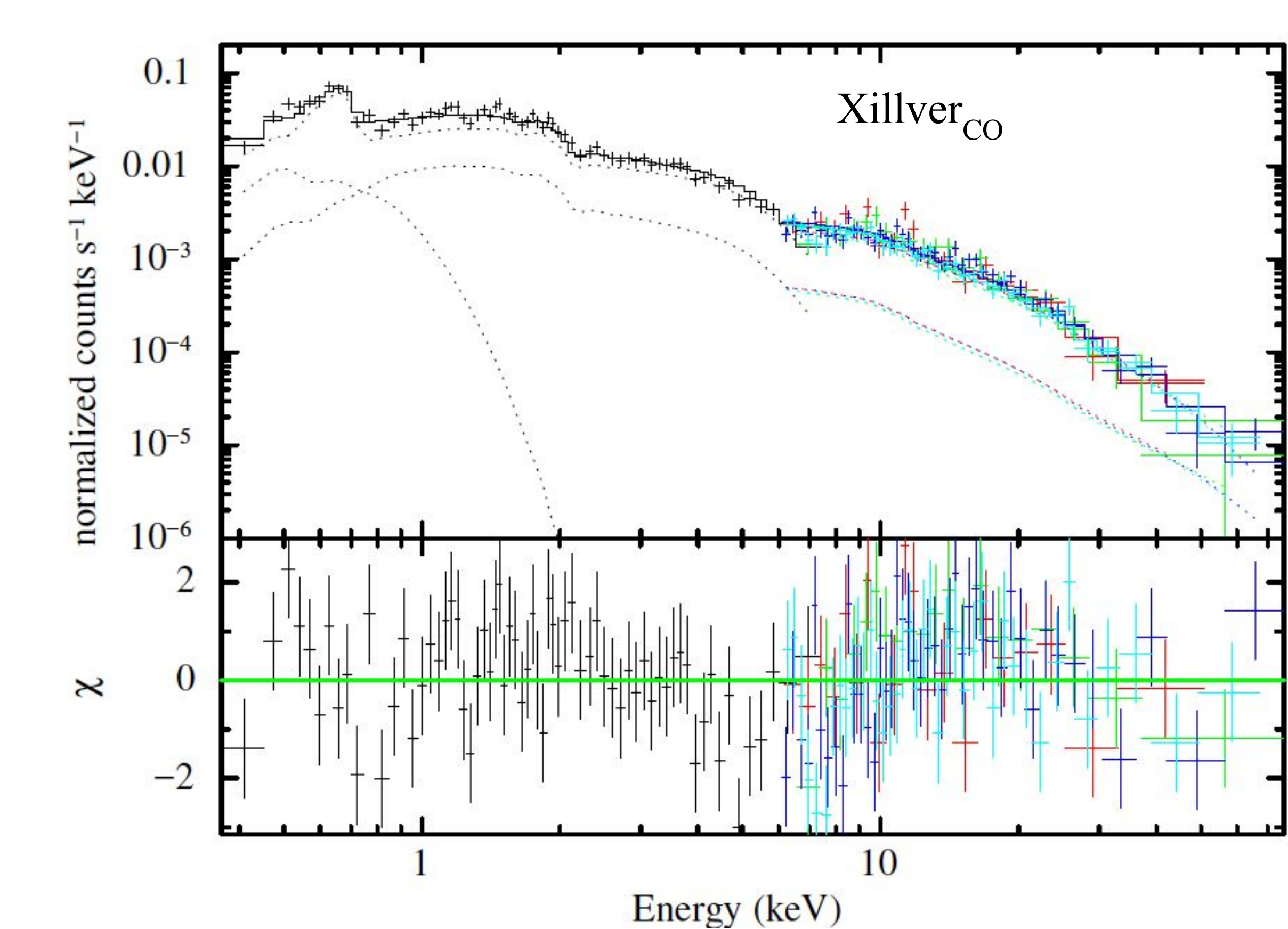}
\includegraphics[scale=0.33]{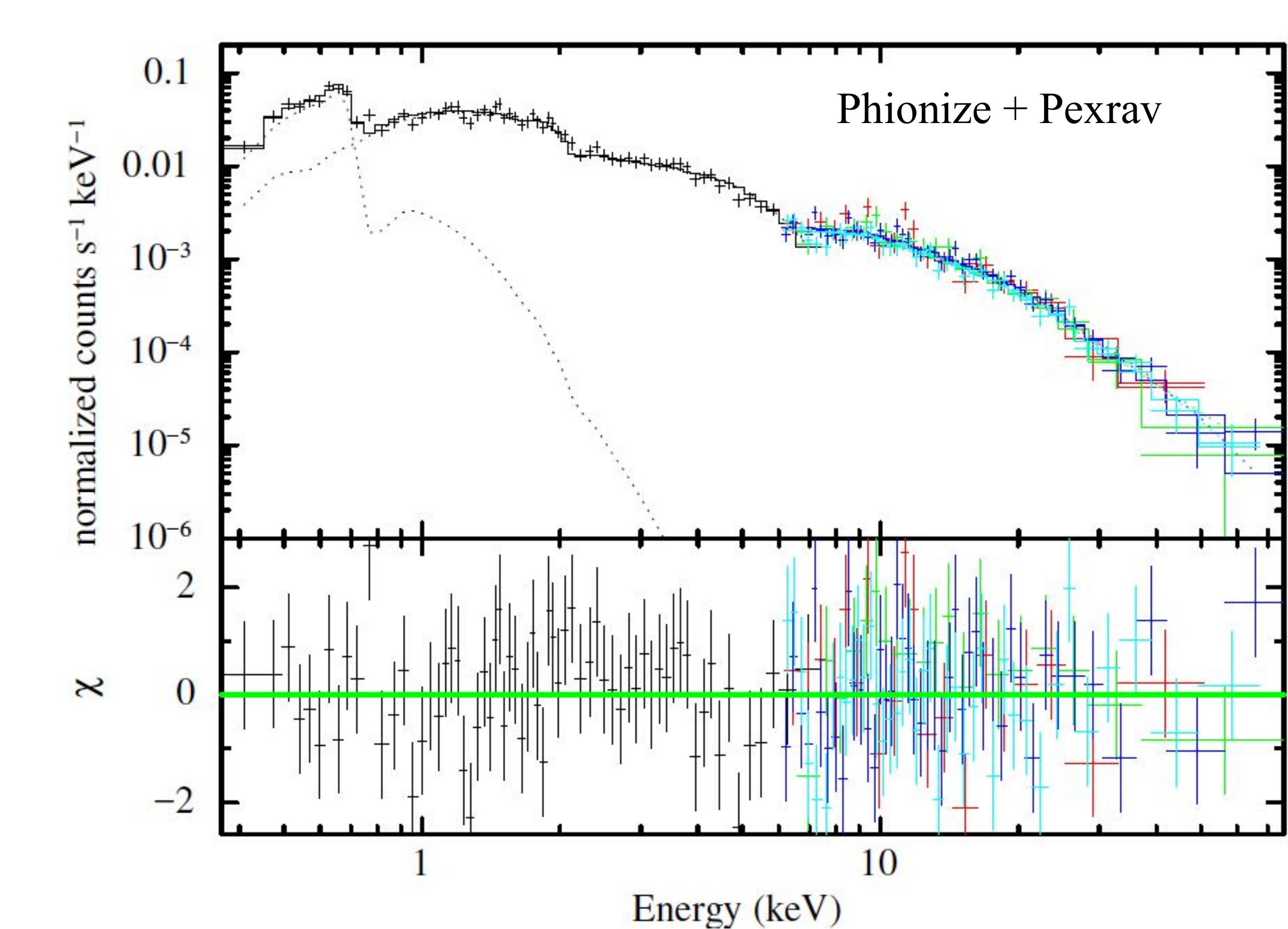}
\caption{X-ray spectrum of 47 Tuc X9 as seen by \chandra\ (black) and \nustar\ (colored, red \& blue from FPMA, green and cyan from FPMB) on February 2015 and models used to fit the spectra. The top panels show two of the fits performed in the 1-79 keV band to address the reflection bump above 10 keV, and the middle and bottom panels show attempts to address the oxygen features below 1 keV. The models represented here are: a cutoff powerlaw (top-left), pexrav (top-right), a single vmekal + pexrav with 47 Tuc cluster abundances (mid-left), two vmekals+pexrav with UCXB abundances (mid-right), UCXB reflection model (bottom-left) and oxygen photo-ionization+pexrav (bottom-right). In the top panels, the reflection bump above 10 keV is clearly addressed by pexrav. Oxygen photo-ionization is the only model properly fitting the oxygen features below 1 keV.}
\label{fig_spec}
\end{center}
\end{figure*}

\begin{figure}
\begin{center}
\includegraphics[scale=0.34]{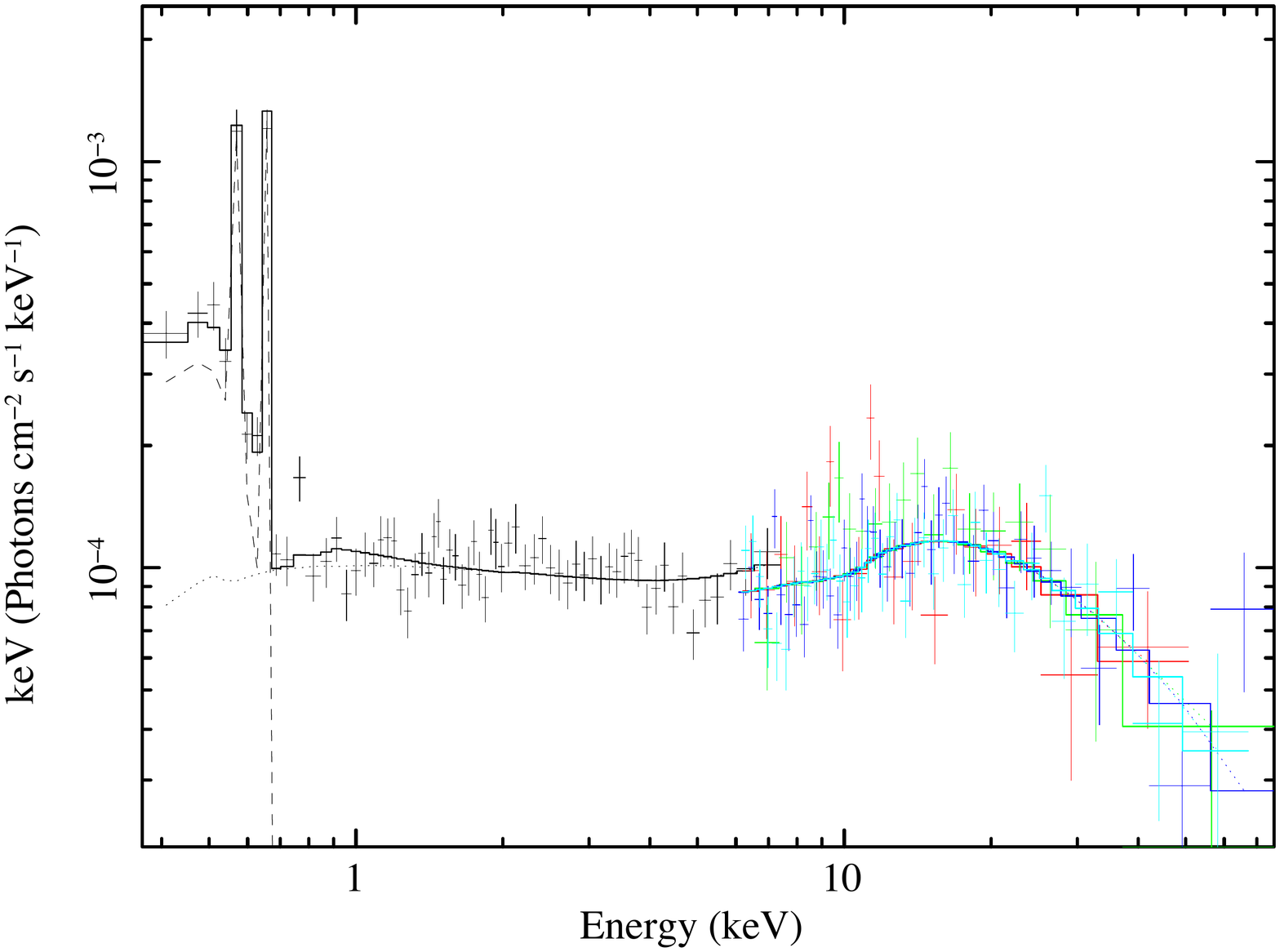}
\caption{Unfolded best-fit model (oxygen photo-ionization+pexrav) describing X-ray spectrum of X9 and the oxygen lines in the system. \chandra\ data and fit shown in black and \nustar\ colored (red \& blue from FPMA, green and cyan from FPMB). The dashed line represents the photo-ionization component, while the dotted line indicates the Pexrav reflection component. The dominant oxygen lines are O VII at 0.57 keV and O VIII at 0.65 keV.}
\label{fig_spec_model}
\end{center}
\end{figure}

We used our XSTAR models to fit the joint \chandra\ \& \nustar\ spectra. To do so, our full model became \texttt{constant*pileup*tbabs*(phoionize+pexrav)}. This model fit the spectra exceptionally well with $\chi_\nu^2 =1.02$ for 181 d.o.f in the case of pure-oxygen and $\chi_\nu^2 =1.03$ for 180 d.o.f in the case of C/O (Table~\ref{tab_spec_soft}), almost completely removing residuals around the oxygen features (Fig.~\ref{fig_spec}, bottom right). Our best fit using the oxygen-only model gives a gas temperature of 0.20$_{-0.10}^{+0.08}$ keV and an ionization parameter ($\log\xi$) value of 0.6$_{-0.1}^{+0.8}$. The best fit with the C/O model gives a similar fit with a gas temperature of 0.24$_{-0.11}^{+0.05}$ keV, ionization parameter ($\log\xi$) of 0.3$_{-0.5}^{+0.1}$ and ratio of carbon/oxygen abundance consistent with zero (unconstrained however). With both models, in our best fits, the Fe abundance in Pexrav is found to be $< 0.3$ of the solar value. \citet{Koliopanos13} suggest that strength of Fe K$\alpha$ line (and thus measured abundance of Iron) should be suppressed in systems with high carbon and oxygen abundances. However, \citet{Madej14} indicate that this is only the case in systems with completely neutral gas. The goodness of fit for the \texttt{phoionize} model clearly indicates present of an ionized region in the system (dominated by oxygen and carbon). However we do not see any evidence for a Fe K$\alpha$ line in the spectrum of X9. This could mean that the accretion disk is nearly neutral and the ionized gas is not part of the disk. It could also mean that the Fe abundance is much lower than solar. This is not surprising as Fe abundance is expected to be low for a 47 Tuc donor star even if not a WD.

We are certain that C exists in the accretion disk, since a strong double-peaked C IV line is detected in the UV spectrum \citep{Knigge08} but no C lines fall into our X-ray fitting range (as they lie below 0.5 keV). We will address constraints on the C/O abundance ratio in a future paper. Using a pure O model gives a good fit, and should not produce significantly different results from the full C/O models (either in goodness of the fit or parameter values) for the spectral fitting range we consider. 

Reproducing our models using the 2.2.1bn24 version of XSTAR yielded an inferior fit ($\chi_\nu^2 =1.2$ compared to 1.02 for 2.2.1bo8, both for 181 d.o.f). This result indicates the importance of bound-bound radiative excitation in production of these lines in the emitting region.

Although our photo-ionized model provides a very good fit, there is a single spectral bin around $\sim 0.8$ keV which shows a significant residual (Fig.~\ref{fig_spec}, bottom right). As this spectral bin is located in a region where the dominant component changes, it is unclear if this residual is due to incomplete modeling of the underlying continuum or caused by a spectral feature. The final model that best fits the spectra is plotted in Fig.~\ref{fig_spec_model}.

The superiority of this fit compared to the collisional models discussed in the previous section is a strong indicator of the extreme overabundance of oxygen and the fact that photo-ionization in this system is not negligible. We discuss the implications of this finding in \S~\ref{sec_ucxb}. 

Our spectral analysis presented above is a first glance at a complex system. We note that in other \chandra\ datasets, due to the strong presence of oxygen features, Mekal with UCXB abundances or our photo-ionized model fits the spectra significantly better than models with solar abundances. We found our model fits most of the \chandra\ spectra reasonably well; however we also noticed significant residual trends in a few data sets. We leave extensive spectral modeling and an in-depth study of the spectral evolution in the system to a later paper (Bahramian et al., in prep.)

\begin{table*}
\centering
\begin{tabular}{@{}lllllll@{}}
\hline
					&	\multicolumn{3}{c}{VMekal}										  & Xillver$_{\text{CO}}$	& \multicolumn{2}{c}{Phoionize}	\\
\cline{2-4} \cline{6-7}					& 	Solar 				&	47 Tuc				&	UCXB 			  &	UCXB					&	Oxygen					&	C/O 	\\
\hline
T1 (keV)			&	0.16$\pm0.01$		& 0.17$_{-0.02}^{+0.01}$& 0.26$\pm0.05$		  &	0.11$\pm0.01$ 			&	0.20$_{-0.10}^{+0.08}$	&	0.24$_{-0.11}^{+0.05}$		\\
T2 (keV)			&	--					&	--					& $<0.3$			  &	--						&	--						&	--		 					\\
$\log(\xi)$			&	--					&	--					& --				  &	--						&   0.6$_{-0.1}^{+0.8}$		&	0.3$_{-0.5}^{+0.1}$			\\
Photon index		&	1.04$\pm0.08$		&	1.07$\pm0.08$		& 1.05$\pm0.08$		  &	1.2$_{-0.2}^{+0.3}$		&	1.12$_{-0.07}^{+0.08}$	&	1.10$_{-0.08}^{+0.07}$		\\
$E_{fold}$	 (keV)	& 	35$_{-7}^{+13}$		&	47$_{-9}^{+14}$		& 38$_{-7}^{+12}$	  &	201$\pm15$				&	51$_{-11}^{+18}$		&	49$_{-11}^{+17}$    		\\
Reflect scale		& 1.3$_{-0.5}^{+3.0}$ 	&	1.1$_{-0.4}^{+1.5}$	& 1.2$_{-0.4}^{+2.0}$ &	--						&	1.1$_{-0.3}^{+1.5}$		&	1.1$_{-0.4}^{+1.4}$			\\
$\theta$ (deg)		&	$<78$				&	$<68$				& $<72$				  &	$<54$					&	$<66$					&	$<46$	 					\\
Norm$_{NU/CX}$		& 0.96$_{-0.08}^{+0.09}$& 0.90$_{-0.08}^{+0.09}$& 0.93$\pm0.08$		  &	1.00$_{-0.07}^{+0.06}$	&	0.89$\pm0.09$			&	0.88$\pm0.09$				\\
$\chi_\nu^2$/d.o.f	& 1.33/183				& 1.23/183				& 1.22/181			  &	1.30/181				&	1.02/181				&	1.03/180 					\\
N.H.P (\%) 			&	0.15				&	1.8					& 2.1				  &	0.43					&	41.6					&	36.0	 					\\
\hline
\end{tabular}
\caption{Best-fit parameter values for our models fitting the entire 0.4-79 keV band spectrum. The VMekal models are VMekal+pexrav models with abundance of elements based on values for solar \citep{Wilms00}, 47 Tuc \citep{Heinke05a,Salaris98,Carney96} and UCXB abundances \citep{Koliopanos13}. Phoionize is the oxygen photo-ionization model produced using XSTAR (see text). T1 and T2 are temperature of each Mekal/VMekal component. Adding a second VMekal component only improved the fit in the case with UCXB abundances. In case of Phoionize, T1 is the temperature of the photo-ionized region. $E_{fold}$ represents the exponential cut-off energy in Pexrav and $\theta$ the inclination angle. For the C/O Phoionize model, we found the best-fit value for carbon to oxygen abundance ratio to be 0, however the value is unconstrained and the uncertainty estimation reaches the model hard limit of 2. Note that the iron abundance parameter in Pexrav is set to the value assumed for VMekal in each case. For Phoionize, this parameter was set free and was found to be $<0.3$ of the solar value in both cases.}
\label{tab_spec_soft}
\end{table*}

\section{Discussion}\label{sec_disc}
\subsection{X9 as an ultra-compact X-ray binary}\label{sec_ucxb}
Based on the X-ray luminosity of X9 ($L_X \sim$ a few $\times 10^{33}$ erg/s) and the inferred mass transfer rate, \citet{Miller-Jones15} suggested the system might be a UCXB with an orbital period of $\sim25$ min. \citet{Knigge08} reported the presence of strong carbon lines in the UV spectrum of X9, indicating the presence of carbon in the accretion flow, and absence of helium lines (hinting at the possibility of a C/O WD donor).

We detect a significant periodic signal in the X-ray light curve with a period of 28.18$\pm0.02$ minutes. This signal is likely the orbital period of the system, indicating that X9 is an ultra-compact X-ray binary. It is worth noting that this periodic signal is mainly present in the 0.5-0.7 keV band dominated by the O VII and VIII lines. This could be foreground column variation due to a disk wind. Since such a foreground column is probably almost pure C/O, there can be a relatively weak wind and it can still provide a significant absorption column, especially in the oxygen edge.

\citet{Heinke05a} showed that in X-rays, X9's spectrum contains strong oxygen lines below 1 keV. As we tried to model the spectrum with various models (collisional gas, reflection in UCXBs, and oxygen photo-ionization; Sec. \ref{sec_spec}), we find the oxygen photo-ionization model to provide a clearly better fit compared to the rest of the models. The superior goodness of this fit indicates the significance of photo-ionization in production of these oxygen lines, and thus points to the presence of overabundant oxygen, possibly in the form of a diffuse region around the inner accretion disk. Clear evidence of photo-ionized oxygen excess provides additional evidence for a C/O WD companion in this system.

Our best-fit model (Section 3.2) indicates the presence of an exponentially cutoff power law, which dominates the emission above 1keV. This component is most likely associated with the inner optically thin accretion flow. However, the fit with the cut-off power law shows correlated residuals in the 10-30 keV band. These residuals largely disappear when a reflection model is convolved with the cut-off power-law. This strongly indicates the presence of reflected emission from an optically thick accretion disk, possibly located further out from the accretor. However, direct X-ray emission from this accretion disk (e.g. a disk blackbody) is not visible, as expected given the low X-ray luminosity of the system. The reflected cut-off power law described all the emission above 1 keV. However, there are strong emission lines below 1 keV that this model fails to describe. These lines are associated with oxygen VII and VIII and seem to be described by photo-ionization (rather than collisional models). Our best-fit model suggests that these lines are produced within an oxygen-rich diffuse gas region possibly embedding the inner parts of the system and irradiated by the accretion flow. The overabundance of oxygen required to explain these lines suggests a hydrogen-poor companion, thus pointing at the ultra-compact nature of the system.

\subsection{Nature of the compact accretor}
Our radio observation of X9 measured flux densities of $27.6\pm7.2$ and $30.7\pm8.9$\,$\mu$Jy\,bm$^{-1}$ at 5.5 and 9.0\,GHz respectively. Comparing with previous observations in 2010 and 2013 \citep{Miller-Jones15},  the 9-GHz radio emission has not changed markedly since the prior observations, although the 5.5 GHz emission appears to have dropped by a factor of 1.5$\pm$0.4. Despite the poor weather and low elevation of X9 for much of the observation, we found no significant evidence for phase decorrelation in the other sources in the field, when comparing their measured fluxes with those determined from the 2010 and 2013 data. 

Is it possible to explain X9 as a CV? Looking simply at the radio, X-ray, and optical fluxes of X9, we see that X9 would be extremely unusual among CVs in its radio flux, its X-ray flux, and its X-ray to optical flux ratio. X9 would have the highest radio flux of any observed CV \citep[e.g., see][]{Russell16,Mooley17,Coppejans16}, and maintains this high radio flux in two widely separated quiescent epochs, in contrast to other CVs, which reach their peak radio brightness during short-lived epochs.

The X-ray flux of X9 is similar to that of the most X-ray luminous known CVs, which reach $L_X$(0.5-10 keV)$\sim5\times10^{33}$ erg/s \citep[e.g., see][]{Stacey11,Bernardini12,Pretorius14}. These systems are all known (or thought) to be intermediate polars, which channel the accretion flow via their magnetic field to the poles, producing shocks above the WD surface \citep[e.g.][]{Aizu73}. This enables strongly magnetic CVs to avoid producing an optically thick boundary layer as in non-magnetic CVs, which shifts the majority of the accretion luminosity into the ultraviolet range \citep{Patterson85}.

However, X-ray luminous intermediate polars have orbital periods of 3.4-10 hours (or more), producing mass transfer rates of $\sim10^{-8}$ \Msun/year \citep[e.g.,][]{Howell01}. For a 1 \Msun WD, this gives $L_{bol}\sim1.5\times10^{35}$ erg/s, allowing them to produce X-ray luminosities of a few $10^{33}$ ergs/s even though most of the released luminosity is not in this energy band \citep[e.g.,][]{Evans09}. We have strong evidence for X9's orbital period being 28 minutes, which indicates a mass transfer rate of $2\times10^{-11}$ \Msun/year for a C/O donor \citep{Deloye03}; for a typical intermediate polar conversion efficiency of mass transfer into X-ray luminosity of 2\%, we would then expect $L_X \sim 6\times10^{30}$ ergs/s, and (implausibly) perfect conversion efficiency would give $L_X=3\times10^{32}$ ergs/s. Increasing the WD mass to the maximal 1.4 \Msun, and assuming perfect conversion efficiency, is required in order to reach $L_X=2\times10^{33}$ ergs/s.

One could also contrive a scenario in which the orbital period is around 10 minutes (producing a mass transfer rate of $10^{-8}$ \Msun/year as in known intermediate polars, using the evolutionary tracks of \citealt{Deloye03}), and the accretion is magnetically channeled to the surface, permitting a large hard X-ray luminosity; but this would require that the true spin and orbital periods of a strongly magnetic WD remain undetected, while a spurious 28-minute period is produced, and an extremely high radio flux is produced. Both of these scenarios appear extremely improbable.

\begin{figure*}
\begin{center}
\includegraphics[scale=0.5]{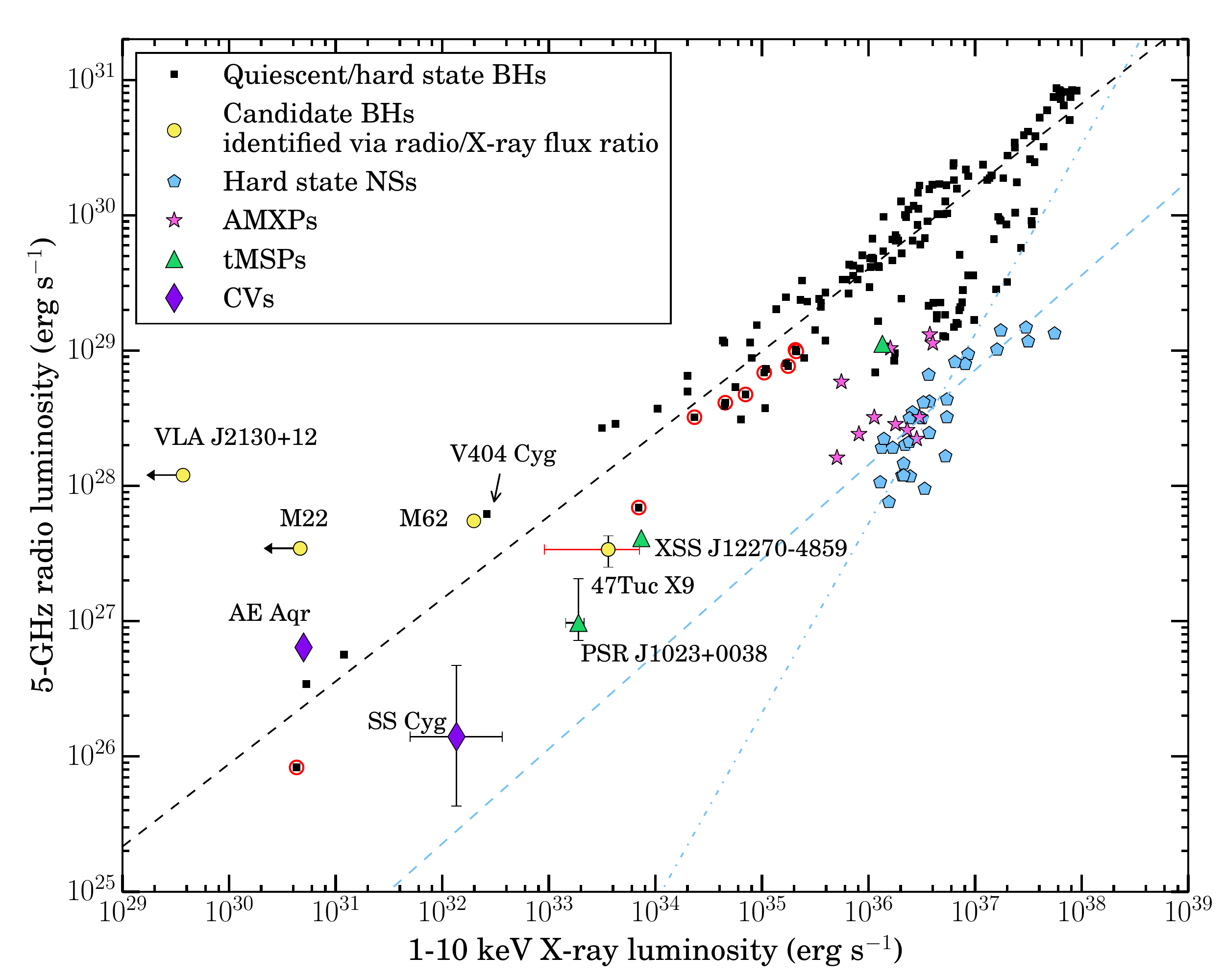}
\caption{Radio/X-ray luminosity correlation for XRBs and stellar mass BHs. Plot adapted from \citet{TetarenkoB16}. X9 falls below the correlation line for BH LMXBs, but consistent with the observed track of XTE J1118+480 (\citealt{Gallo14}, denoted by red circles around black squares in this plot). Note that the uncertainty on X-ray luminosity of 47 Tuc X9 is relatively small ($\sim3\times10^{31}$ erg/s) and the red bar in this plot just demonstrates the scale of X-ray variability. The data point for SS Cyg and AE Aqr show their radio and X-ray luminosities during their flaring states \citep{Russell16}.}
\label{fig_lrlx}
\end{center}
\end{figure*}

With our simultaneous X-ray and radio measurements of X9 from February 2015, we plot X9 on the L$_R$-L$_X$ plane (Fig.~\ref{fig_lrlx}). Compared to \citet{Miller-Jones15}, we find a lower 5.5 GHz luminosity, but a higher X-ray luminosity.\footnote{Note that the observations reported in \citet{Miller-Jones15} were not simultaneous.}  This places X9 within the scatter of the BH population (and near the track followed by XTE J1118+480, \citealt{Gallo14}), but closer to the tMSPs.

The position of X9 on the L$_R$-L$_X$ plot rules out a CV but cannot discriminate between a (tMSP) NS or a BH as possibilities for the accreting object. There have only been 10 quiescent BH LMXBs with detailed X-ray analyses \citep{Plotkin13}, and only $\sim$3-4 quiescent BH LMXBs with deep radio/X-ray studies \citep[e.g.,][]{Plotkin16}. The X-ray properties of X9 are significantly different from those of any known tMSP or BH LMXB. X9 has a significantly harder continuum spectrum than either tMSPs \citep[photon indices 1.5-1.7,][]{deMartino10,Linares14,Bogdanov15,Bogdanov15b} or BH LMXBs in quiescence \citep[photon indices in the range 1.5-2.3,][]{Plotkin13,Wijnands15}. It is worth noting that tMSPs in rotation-powered state can show a hard spectrum \citep[with photon index $\sim1.1$,][]{deMartino15}. However X9 is too bright (both in radio and X-rays) to be in this state. Fitting X9's 0.5-10 keV spectrum (in subarray mode to eliminate pile-up) gives a photon index of 1.06$\pm$0.24 \citep{Miller-Jones15}. We might tentatively speculate that X9's unusual continuum spectral shape might be related to the unusual chemical composition of its donor. Similarly, the 6.8-day order-of-magnitude variation in X9's X-ray flux is not similar to any behaviour seen in either tMSPs or BH LMXBs.  Detection or ruling out pulsations would be strong evidence, but the pulsed fraction limit of $<19$\% that we were able to set from the Chandra HRC data alone cannot rule out a tMSP nature, since for some tMSPs the coherent X-ray pulsations have pulsed fractions down to 8\% \citep{Archibald15,Papitto15}. Detection of a sharp change in the Fermi gamma-ray flux from 47 Tuc coincident with a state change in X9 would also be a strong signal that X9 is a tMSP \citep{Stappers14,Johnson15}. No such change is visible in the long-term Fermi lightcurves from 47 Tuc, but this is not strongly constraining considering the presence of $>$20 gamma-ray-emitting MSPs in the cluster \citep[e.g. ][]{Camilo00,Freire03,Abdo09c,Pan16,Ridolfi16}.

An inspection of its variability properties shows X9 to behave somewhat differently to the known tMSPs in both the radio and X-ray bands, although we note that the samples of well-studied tMSP and quiescent BH sources are both small. First, the three verified tMSPs (and a candidate tMSP) show rapid (hundreds to thousands of seconds) switching between two relatively stable luminosity levels around $10^{33}$ and $10^{34}$ erg/s \citep{Papitto13,deMartino13,Linares14,Tendulkar14,Bogdanov15,Bogdanov15b}. However, \citet{Miller-Jones15} were able to show that the available X9 X-ray data does not show sudden dips to a ``low mode'', or rapid flaring states, as seen in the X-ray lightcurves of the three known tMSPs \citep{Bogdanov15,deMartino13,Linares14}; of order 100 detectable drops to the ``low mode'' should have been detected in the X9 Chandra data, should it show a similar distribution of ``low mode'' dip frequency with dip length (e.g. dips longer than 200 s on average every 3 ks) as PSR J1023+0038 or XSS J12270-4859. 

Second, X9's radio flux appears quite stable (varying by less than a factor of 3) between different observing epochs (10-hour runs in 2010 and 2013, and 12 hours in 2015). This is in contrast to PSR J1023+0038, which shows dramatic variations of up to a factor of 100, varying by a factor of 10 on as short a timescale as 30 minutes \citep{Deller15}. Such strong variations in the radio have not been seen from BH LMXBs in quiescence \citep[see, e.g.][]{Gallo14,Hynes04,Miller-Jones08}. Thus, X9 would appear more similar to BH LMXBs in both its X-ray and radio variability properties than to tMSPs.  However with only three tMSPs confirmed so far, we cannot be certain that all tMSPs will show the same patterns of X-ray and radio variability as the known sources, so we cannot rule out a tMSP as a possibility, but the evidence suggests a moderate preference for a BH nature of the accretor.

An alternative scenario is that we are viewing the system at a very high inclination angle, where the outer disk obscures the inner parts of the accretion disk, so that the observed X-ray flux is significantly lower ($\sim$2-3 orders of magnitude) than the true flux. We found clear evidence for reflected emission in the X-ray spectrum of X9. To our knowledge, this is the lowest-luminosity X-ray binary (non-CV) system to show reflection. Clear detection of reflection from a faint ($<10^{34}$ erg/s) system and the ionized emission possibly associated with an accretion disk corona in the system, dominating the continuum below 1 keV, suggests that the inner accretion disk might be heavily obscured and thus the true X-ray luminosity could be higher. In this scenario, a NS nature for the compact object would be preferred by the radio/X-ray flux ratio. We note that this case requires a very high inclination angle which is inconsistent with the constraints indicated by the reflection models (Table \ref{tab_spec_hard}). However, due to assumptions of the abundances of elements in these models (which do not consider hydrogen-deficient gas), these constraints could be inaccurate.  In this scenario, the system is persistently bright, at an X-ray luminosity of $\sim10^{36}$ erg/s, which is indeed consistent with the time-averaged mass transfer rate for a NS X-ray binary in a 28-minute orbital period \citep{Deloye03,vanHaaften12b}. However, we find no evidence for eclipses, while partial eclipses of the scattering region are invariably seen in such accretion disk corona sources \citep[e.g.,][]{White82,Frank87}. The variability seen at 28 minutes takes far too much of the orbit to be an eclipse from a companion. Thus, we find the scenario of a very high inclination angle to be less plausible than the tMSP or BH LMXB models discussed above.

\subsection{Origin of the super-orbital periodicity}\label{sec_disc_suporb}
As discussed by \citet{Miller-Jones15}, the relatively high quiescent X-ray luminosity of X9 suggests a relatively high mass-transfer rate, only slightly below the rate needed to maintain the disk in a persistently bright state. Such a high mass-transfer rate could be reached through standard binary evolution of a low-mass companion either at large orbital periods (such as the 6.5 day orbital period of V404 Cyg, \citealt{Casares92}), or short orbital periods around 25 minutes. However, as discussed by \citet{Miller-Jones15}, any orbital periods larger than 3 hours are clearly ruled out based on the limits on optical emission from the companion star. This paper presents robust evidence for a WD companion in the system (strong oxygen emission lines, and a periodic signal at 28.18 minutes). Thus, the 6.8 day periodicity, if real, must have a super-orbital origin.

Super-orbital modulations have been observed in multiple XRBs (e.g., see \citealt{Charles08} for a review). It is thought that these modulations can be produced by mechanisms like precession, tilting or warping of the accretion disk. 

If this super-orbital period is due to superhump precession, using the orbital period of $P_{\text{orb}}=28.2$ min and taking a precession period of $P_{\text{pre}}=6.8$ days, we can estimate the mass ratio $q$, using the relation $q = 0.114+3.97\times(\epsilon-0.025)$, where $\epsilon$ is the superhump excess and can be approximated by $P_{\text{orb}}/P_{\text{pre}}$ \citep{Knigge06}. Using this relation we find $q\sim0.026$. A UCXB with an orbital period of 28.2 min indicates a donor white dwarf mass of $\sim 0.02 M_\odot$ \citep[e.g.,][]{Rappaport87} and thus suggests an accretor mass of $\sim1 M_\odot$. We note that there are many uncertainties in this calculation; the superhump relations are derived empirically based on CVs, and are not validated for such low $q$ values.  If correct, this calculation would argue for an NS accretor; however, since we are not certain the 6.8 day periodicity is indeed periodic, that it is due to precession, or that the precession is necessarily due to the superhump instability, we cannot consider it strong evidence on the nature of the accretor. However, this calculation does indicate that a 6.8 day precession period is physically reasonable for a system with an orbital period of 28.2 min.

Among the systems showing super-orbital modulation, 4U 1916-053 is a persistent UCXB ($\sim$ 1000 times brighter than X9) with an orbital period of 50 min \citep{Walter82,Nelemans10a}. Strong helium and nitrogen lines in the spectrum indicate a He WD companion \citep{Nelemans06}, and X-ray bursts indicate an NS accretor \citep{Becker77}. 
The short orbital period (1.8 hour) system UW CrB (MS 1603+2600) shows a super-orbital modulation of period $\sim$5 days \citep{Mason08,Hakala09}, presumably related to the ``superhump'' phenomenon \citep{Patterson05}. \citet{Charles08} suggest that the 5-day modulation in this system\footnote{\citet{Charles08} incorrectly attribute UW CrB's superorbital period to 4U 1916-053.} is caused by precession and/or warping of the accretion disk. Given the similarities between UW CrB and 47 Tuc X9, it is possible that similar mechanisms are responsible for the 6.8 day modulation observed in X9.

\section*{Acknowledgements}
AB thanks E.W.~Koch for help with computational aspects of this project, K.A.~Arnaud for helpful discussion on spectral analysis and B.E.~Tetarenko for help with the Radio/X-ray luminosity plot. The authors thank F.A. Harrison for granting \nustar\ director's discretionary time for these observations and J.~Tomsick for assisting in coordination of \nustar\ observations. JCAMJ is the recipient of an Australian Research Council Future Fellowship (FT 140101082). JS acknowledges support of NSF grant AST-1308124 and a Packard Fellowship. COH and GRS acknowledge support from NSERC Discovery Grants, and COH also acknowledges support from a Humboldt Fellowship. This work was funded in part by NASA \textit{Chandra} grant GO4-15029A awarded through Columbia University  and issued by the
\textit{Chandra} X-ray Observatory Center, which is operated by the Smithsonian Astrophysical Observatory for and on behalf of NASA under contract NAS8-03060.

The scientific results reported in this article are based on observations made by the \chandra\ X-ray Observatory, \nustar\ observatory and Australia Telescope Compact Array, and archival data obtained from \chandra\ and \Swift\ data archives. The Australia Telescope Compact Array is part of the Australia Telescope National Facility which is funded by the Australian Government for operation as a National Facility managed by CSIRO.

We acknowledge use of the following packages/softwares for analysis: the NuSTAR Data Analysis Software (NuSTARDAS) jointly developed by the ASI Science Data Center (ASDC, Italy) and the California Institute of Technology (Caltech, USA), CIAO; provided by the Chandra X-ray Center (CXC), HEASoft; provided by the High Energy Astrophysics Science Archive Research Center (HEASARC), a service of the Astrophysics Science Division at NASA/GSFC and of the Smithsonian Astrophysical Observatory's High Energy Astrophysics Division, PVM\_XSTAR; developed at the MIT Kavli Institute for Astrophysics and funded in part by NASA and the Smithsonian Institution, through the AISRP grant NNG05GC23G and Smithsonian Astrophysical Observatory contract SV3-73016, Astropy \citep{Robitaille13}, AstroML \citep{Vanderplas12}, Gatspy \citep{Vanderplas15} and Matplotlib \citep{Hunter07}. Our simulations were performed using Cybera cloud computing resources (\url{http://www.cybera.ca/}). We acknowledge extensive use of NASA's Astrophysics Data System and Arxiv.

\bibliographystyle{mnras}
\bibliography{ref_list} % if your bibtex file is called example.bib

\bsp	% typesetting comment
\label{lastpage}
\end{document}